\documentclass[12pt]{iopart}
\usepackage[english]{babel}

\usepackage[letterpaper,top=2cm,bottom=2cm,left=3cm,right=3cm,marginparwidth=1.75cm]{geometry}

\expandafter\let\csname equation*\endcsname\relax
\expandafter\let\csname endequation*\endcsname\relax

\usepackage{amsmath}
\usepackage{amsfonts}
\usepackage{graphicx}
\usepackage{subcaption}
\usepackage{enumitem}
\usepackage{float}
\usepackage{comment}
\usepackage[sorting=none]{biblatex}
\usepackage[colorlinks=true, allcolors=black]{hyperref}

\begin{document}
\title[Anomaly detection with spiking neural networks for LHC physics]{Anomaly detection with spiking neural networks for LHC physics}

\author{Barry M. Dillon, Jim Harkin, Aqib Javed}

\address{ISRC, Ulster University, Derry, BT48 7JL, Northern Ireland}

\ead{b.dillon@ulster.ac.uk}

\vspace{1pt}

\begin{abstract}
\noindent Anomaly detection offers a promising strategy for discovering new physics at the Large Hadron Collider (LHC).
This paper investigates AutoEncoders built using neuromorphic Spiking Neural Networks (SNNs) for this purpose.
One key application is at the trigger level, where anomaly detection tools could capture signals that would otherwise be discarded by conventional selection cuts.
These systems must operate under strict latency and computational constraints. 
SNNs are inherently well-suited for low-latency, low-memory, real-time inference, particularly on Field-Programmable Gate Arrays (FPGAs). 
Further gains are expected with the rapid progress in dedicated neuromorphic hardware development. 
Using the CMS ADC2021 dataset, we design and evaluate a simple SNN AutoEncoder architecture. 
Our results show that the SNN AutoEncoders are competitive with conventional AutoEncoders for LHC anomaly detection across all signal models. 
\end{abstract}
\vspace{4pt}

%\tableofcontents
%\maketitle

\section{Introduction}

\noindent
A primary goal of the Large Hadron Collider (LHC) is to search for new physics beyond the Standard Model. 
The ATLAS and CMS collaborations have conducted extensive searches using traditional analysis techniques, but no compelling evidence of new particles has emerged.
These conventional approaches typically define a signal model and search for a limited set of associated signatures. 
However, given the many possible new physics signatures, more general anomaly detection methods are increasingly attractive. 
These methods do not target specific models; they instead search for events in the data that deviate from the expected backgrounds.
Both ATLAS and CMS have previously explored such approaches \cite{Aaboud:2018ufy,2021:cmsmusic}, but no significant deviations from backgrounds were observed. 
A major challenge in these strategies is the high dimensionality of the data. 
Histogram-based methods struggle with this due to the curse of dimensionality.
Modern machine-learning techniques using neural networks offer a promising alternative for tackling this complexity.

Within the realm of neural networks, AutoEncoders (AEs) are a powerful tool for anomaly detection.
They have shown the ability to identify anomalous jets in LHC data with sparse high-dimensional feature spaces, $\!\sim\!\mathcal{O}(1600)$ \cite{Hajer:2018kqm,Heimel:2018mkt,Farina:2018fyg}.
Complementary to this are more interpretable approaches, for example in \cite{Dillon:2025dxr} the authors use theory-informed reinforcement learning to map events to partonic descriptions of background processes, using the background matrix-element as the anomaly score.
Probabilistic models offer another approach \cite{Dillon:2019cqt,Dillon:2020quc,Dillon:2021nxw}, capable of identifying patterns of new physics in large feature spaces for signal fractions as low as a few percent.
Semi-supervised methods such as Classification Without Labels (CWoLa) also have strong advantages \cite{Metodiev:2017vrx},
incorporating background-estimation techniques closely aligned with traditional strategies like the bump hunt \cite{Collins:2018epr,Collins:2019jip,Kasieczka:2021xcg,Nachman:2020lpy,Hallin_2022,Raine:2022hht}.

Significant progress has been made in developing AutoEncoders for anomaly detection in particle physics.
These efforts have addressed robustness \cite{Roy:2019jae,Blance:2019ibf,Finke:2021sdf,Bradshaw:2022qev,Khosa_2023,Banda:2025nrv}, explored variational AutoEncoders \cite{Cerri:2018anq,Pol:2020weg,Jawahar:2021vyu,Dillon:2021nxw,Fraser:2021lxm,Buss_2023,Cheng_2023}, permutation-invariant architectures \cite{Atkinson:2021nlt,Atkinson:2022uzb,Ostdiek_2022}, and quantum AutoEncoders \cite{Ngairangbam:2021yma,Araz:2024lsl,Bal:2025ydm}.
AutoEncoders have also been applied to semi-visible jets \cite{Canelli:2021aps,Dillon:2022mkq,Faucett:2022zie,Chhibra:2023tyf}, where the signals are subtle and difficult to detect in a model-independent way.
The role of symmetry in particle physics data has inspired Lorentz-equivariant architectures such as \cite{Hao_2023}, and self-supervised methods that incorporate invariances into the data representations \cite{Dillon:2021gag,Dillon:2022tmm,Dillon:2023zac,Favaro:2023xdl,Matos:2024ggs}.
These approaches are typically designed for offline analyses using data recorded to disk.
This raises the risk that new physics signals might be missed at the trigger-level and not recorded.

Real-time anomaly detection at the trigger-level presents a compelling opportunity for new physics searches with AutoEncoders \cite{Govorkova:2021hqu}.
The LHC collides protons at a rate of $40$MHz, but only a small fraction can be recorded for offline analysis ($\sim1000$/s).
CMS manages this using a two-tier trigger system.
The Level-1 (L1) trigger uses low-level calorimeter data to reduce the rate to $100$ kHz (latency $\sim 4\mu$s).
The High-Level Trigger (HLT) then applies full event reconstruction and further reduces the rate down to $1$ kHz.
While the triggers are optimized to select events relevant for Standard Model studies and anticipated new physics topologies, there remains a risk that unexpected new physics signatures are being discarded.
Neural network based trigger systems such as AutoEncoders could help mitigate this by identifying additional interesting events at the L1 trigger to be recorded \cite{Govorkova:2021utb,Mikuni:2021nwn,Kosters:2022amb,Jia:2024ysq}.
These neural networks must run at very low-latency when deployed on specialized hardware.

Current methods for real-time anomaly detection typically rely on deep neural networks deployed on FPGAs (Field Programmable Gate Arrays) for low-latency inference.
Techniques such as network pruning and weight quantization \cite{han2016deepcompressioncompressingdeep} are used to meet latency requirements.
Pruning removes unnecessary parameters from the network while attempting to preserve performance, 
and quantization reduces the floating-point precision of the weights to accelerate the multiply-and-accumulate operations on the FPGA.
Quantization can be performed post-training, or during training using Quantization Aware Training (QAT).
The HLS4ML package has been developed to translate machine-learning models to firmware implementations that can be used on FPGAs \cite{fastml_hls4ml,Duarte:2018ite}.
Recently, BitNet-based architectures have also been explored in particle physics and may offer improved efficiency \cite{Krause:2025qnl}, though they remain untested in anomaly detection tasks.
More broadly, low-latency machine-learning (FastML) is gaining traction across scientific domains \cite{Deiana:2021niw}, and a collection of FastML science benchmarks has been compiled in \cite{Duarte:2022hdp}.

Spiking Neural Networks (SNNs) offer an alternative approach to low-latency and memory-efficient deep learning \cite{TAVANAEI201947,oconnor2016deepspikingnetworks}.
While architecturally similar to conventional neural networks, SNNs transmit information using discrete spikes rather than continuous floating-point numbers.
Unlike regular neural networks, SNNs naturally process information in steps.
There are many applications of SNNs to time-series data where each step corresponds to a time-step.
For static forms of data these steps can be used to represent other features in the data.
First introduced in the 1970s \cite{ANDERSON1972197}, SNNs are receiving renewed attention due to their potential for ultra-low-latency and low-power inference.
They can be deployed on FPGAs for efficient inference, but can benefit significantly from dedicated neuromorphic hardware.
Recent years have seen significant advances in the development of application-driven neuromorphic hardware with IBM TrueNorth \cite{2014Sci...345..668M}, Intel Loihi \cite{8259423,10448003}, and SpiNNaker \cite{6750072}.

SNNs have already attracted interest within the physics community.
They have been shown to perform well in classifying time-series data from the MINERvA experiment at Fermilab  \cite{Schuman:2017kdp,Aliaga_2014}, performing comparably to Convolutional Neural Networks (CNNs) while using significantly fewer resources.
For LHC physics SNNs have been studied for filtering sensor data \cite{Kulkarni:2023lpb}, unsupervised particle tracking \cite{Coradin:2025ees}, and jet-tagging \cite{bartlomiej_borzyszkowski_2020_3755310}.

This paper presents a first exploration of SNNs for anomaly detection at the LHC. 
Our results demonstrate that despite their limited computational capacity, SNN-based AutoEncoders (SNN-AEs) perform competetively with AEs constructed from DNN layers. 
Moreover, the SNN-AEs appear more stable across variations in architecture and training dataset size. 
Sec. \ref{sec:snns} introduces the basics of SNNs, including how information passes through the network, how they are trained, and what leads to the low-latency and computational efficiency.
Sec. \ref{sec:data} introduces the CMS anomaly detection challenge dataset \cite{Govorkova:2021hqu} that was used to test the performance of the SNNs.
In Sec. \ref{sec:snnaes} we define the SNN-AE architecture and present a comparison between the SNN-AE and the conventional AE on the CMS ADC dataset.
Finally, Sec. \ref{sec:conclusions} summarizes the results and outlines directions for future work.

\section{Spiking Neural Networks}
\label{sec:snns}

Spiking Neural Networks (SNNs) share a similar architecture with conventional Deep Neural Networks (DNNs), but differ fundamentally in how they process and transmit information through the network.
SNNs are designed for fast and efficient inference by incorporating neuro-inspired mechanisms for the propagation of information through networks.
Each neuron in a layer of a DNN is connected to each neuron in the subsequent layer.
Information is propagated through the network via large matrices of floating-point numbers.
SNNs are also fully connected, but only propagate information between neurons via one of two values: a $1$ (spike) or a $0$ (no spike).
The neurons in a conventional DNN also have no internal state; they are memoryless, and the output of the network is obtained in a single forward pass through the neural network. 
Neurons in SNNs, on the other hand, do have an internal state, and the output of a network is obtained after several steps in which information is passed through the network.

\subsection*{Spiking neurons}

A neuron in a conventional DNN computes
\begin{equation}
\label{eq:nnact}
y=f(wx+b)
\end{equation}
where $x$ is the input to the neuron, $y$ is the neuron output, $f$ is the activation function, and $(w,b)$ are the learnable weight and bias terms.
A spiking neuron, on the other hand, is governed by
\begin{align}
    y_t &= \begin{cases}
0 & \text{if } u_t < u_{\text{thresh}} \\
1 & \text{if } u_t \geq u_{\text{thresh}}
\end{cases} \\
    u_{t+1} &= \beta u_t + w x_{t+1} - \beta y_t u_{\text{thresh}} + b.
\end{align}
The neuron generates a spike at step $t$ when the neuron potential $u_t$ exceeds the neurons threshold potential $u_{\text{thresh}}$.
The neuron potential at step $t\!+\!1$ gets contributions from the neuron input at that step, $x_{t+1}$, and the previous neuron potential $u_t$, therefore it builds up with each input at each step.
The contributions from previous steps are controlled by the decay factor $\beta$, and once the neuron spikes ($y_t\!=\!1$) the neuron potential resets.
So the spiking neuron has a state that persists and updates between each step.
Both types of neuron have a learnable weights and biases, but their sources of non-linearity differ:
DNNs use a differentiable non-linear activation function $f$, while SNNs have a discontinuous spike-reset mechanism.
The spiking neuron above is an example of the well-known Leaky Integrate and Fire (LIF) neuron.
The $\beta$ and $u_{\text{thresh}}$ are hyper-parameters of the LIF neuron that we can tune for our particular use-case.

\subsection*{Layers of spiking neurons}

Neural networks constructed from spiking neurons mirror the structure of conventional networks.
For a single layer neural network $y:\mathbb{R}^M\rightarrow\mathbb{R}^N$, the network computes
\begin{equation}
    \vec{y}(\vec{x}) = f\left( W \vec{x} + \vec{b}\right)
\end{equation}
where $\vec{x}$ is the input vector of length $M$, $W$ is a matrix of dimension $N\times M$, and $\vec{b}$ is a vector of length $N$.
The output vector $y$ has dimension $N$, the parameters of $W$ and $\vec{b}$ are learnable, and the activation function is applied element-wise on $W \vec{x} + \vec{b}$.
For the spiking neuron layer, we write
\begin{align}
    \vec{y}_t &=  \Theta\left( \vec{u}_t - u_{\text{thresh}} \right) \\
    \vec{u}_{t+1} &= \beta \vec{u}_t + W \vec{x}_{t+1} - \beta \vec{y}_t u_{\text{thresh}} + \vec{b}
\end{align}
where the neuron potentials $\vec{u}_t$ are vectors of length $N$, $W$ again plays the role of the learnable weight matrix with dimension $N\times M$, and the bias term has dimension $N$.
The Heaviside step function $\Theta$ is used here to indicate that only neurons whose potential exceeds the threshold generate a spike.
When we build deeper networks with more than one layer, the outputs ($y$ or $y_t$) of the first layer become the inputs ($x$ or $x_t$) of the next, and so on.
All inter-layer communication occurs via binary spikes, except at the input layer, which can receive either a series spikes or continuous floating-point numbers fed directly to the neuron potentials.

\subsection*{The forward-pass}

The forward-pass in a conventional network involves a single evaluation; we pass the data through the network once.
In a complete forward-pass through an SNN the data is passed through the network $T$ times, i.e. $T$ steps.
The number of steps $T$ is a hyper-parameter that we can optimize.
At each step the neuron potentials change and some neurons spike.
Upon completion of all steps the neuron potentials are reset.

The data we are considering here is static; it does not have a natural representation as a series of steps, like time-series data.
We could find a way to encode this data as a series, and input it to the SNN in this way.
Instead, we adopt a simpler strategy: the same continuous input is fed to the input neuron potentials at each step. 
The neurons accumulate inputs until they fire, and the spikes propagate information through the network.
On the output layer of the SNN we increase the threshold potential $u_{\text{thresh}}$ such that the output layer neurons will not spike.
At the end of the $T$ steps we can then read off the neuron potentials that have accumulated on the output layer, and treat these as the network output.
These choices mean our SNN workflow more closely resembles the conventional neural network workflow, thus facilitating a like-for-like comparison.

\subsection*{Backpropagation}

A key feature of DNNs is the differentiability of the activation function $f(wx\!+\!b)$ in Eq.~\ref{eq:nnact} enabling the optimization of the weight and bias terms via backpropagation.
However the activation mechanism in the spiking neurons is clearly not differentiable.
Surrogate gradients use continuous differentiable approximations of the activation mechanism in spiking neurons (such as the arctan function) to approximate the gradients used to optimize their weights \cite{oconnor2016deepspikingnetworks,8891809}.
Many streamlined tools already exist for the optimization of DNNs, such as \texttt{pytorch} \cite{Ansel_PyTorch_2_Faster_2024}.
The \texttt{snntorch} \cite{eshraghian2021training} package has been developed on top of \texttt{pytorch}, making use of the built-in autograd functionality to optimize the weights in SNNs using surrogate gradients.
All results in this paper were arrived at using these tools.

\subsection*{The efficiency of SNNs}
SNNs offer substantial benefits in latency and efficiency compared to standard DNNs, particularly when deployed on FPGAs or neuromorphic hardware \cite{LI2022102765}.
The key difference is in how the data is processed within the networks.
DNNs rely on multiply-accumulate (MAC) operations, while SNNs can be implemented using simpler operations such as multiplex-accumulate (MUX) or conditional-accumulate \cite{KARAMIMANESH2025107256}.

In a DNN, each neuron processes all inputs on every cycle, regardless of their relevance to the computation.
Even with pruning and quantization, the resulting DNNs still perform many redundant operations unless they are very carefully optimized.
In contrast, an SNN only processes information when a neuron spikes, drastically reducing the computational overhead \cite{Yao2024}.
This is called event-based processing.
Because the spikes are binary, the SNN is free of expensive matrix-multiplication operations, the MAC operation reduces to a conditional accumulate operation and is much more efficient \cite{plagwitz2023spikespikequantitativecomparison}.
FPGAs and neuromorphic hardware can be designed to take advantage of these more efficient operations.
Neuromorphic hardware in-particular uses a combination of local memory, asynchronous computation, and sparse spike-based communication to boost the performance of SNN architectures \cite{rathi2023exploring}.
As both hardware and training techniques improve, the efficiency advantage of SNNs is expected to grow.

\section{CMS dataset}
\label{sec:data}
This paper aims to evaluate the performance of SNNs in anomaly detection tasks for the LHC, with a particular focus on online, real-time applications at the trigger level.
A prime example of this is the real-time trigger system at CMS.
For this study we use the CMS Anomaly Detection Challenge (ADC2021) dataset \cite{Govorkova:2021hqu}.
The simulated events are filtered by requiring at least one electron with $p_T\!>\!23$ GeV and $|\eta|\!<\!3$, or one muon with $p_T\!>\!23$ GeV and $|\eta|\!<\!2.1$.
The following high-level information is then retained to determine anomaly scores for each event:
\begin{itemize}
\itemsep0pt
    \item up to $10$ jets - $p_T\!>\!30$ GeV \& $|\eta|\!<\!4$
    \item up to $4$ muons - $p_T\!>\!3$ GeV \& $|\eta|\!<\!2.1$
    \item up to $4$ electrons - $p_T\!>\!3$ GeV \& $|\eta|\!<\!3$
    \item the missing transverse energy (MET).
\end{itemize}
Although each object (except MET) includes $(p_t,\eta,\phi)$, we only use the $p_T$ values in this analysis.
This results in a $19$-dimensional vector describing each event ($18$ $p_T$'s from jets, electrons, and muons, and the MET).
The information is ordered as described above, with the first ten entries reserved for jets, the next four for muons, and the next four for electrons, with MET occupying the final entry. Within each category, the jets and leptons are ordered by $p_T$ in descending order; for example, entry $0$ will contain the highest $p_T$ jet, and entry $4$ the highest $p_T$ muon, if the event contains jets and muons.
If an event has fewer than the maximum recorded jets or leptons, then those corresponding entries contain zeros.

\subsection*{The background dataset}
\begin{figure}[H]
    \includegraphics[width=\textwidth]{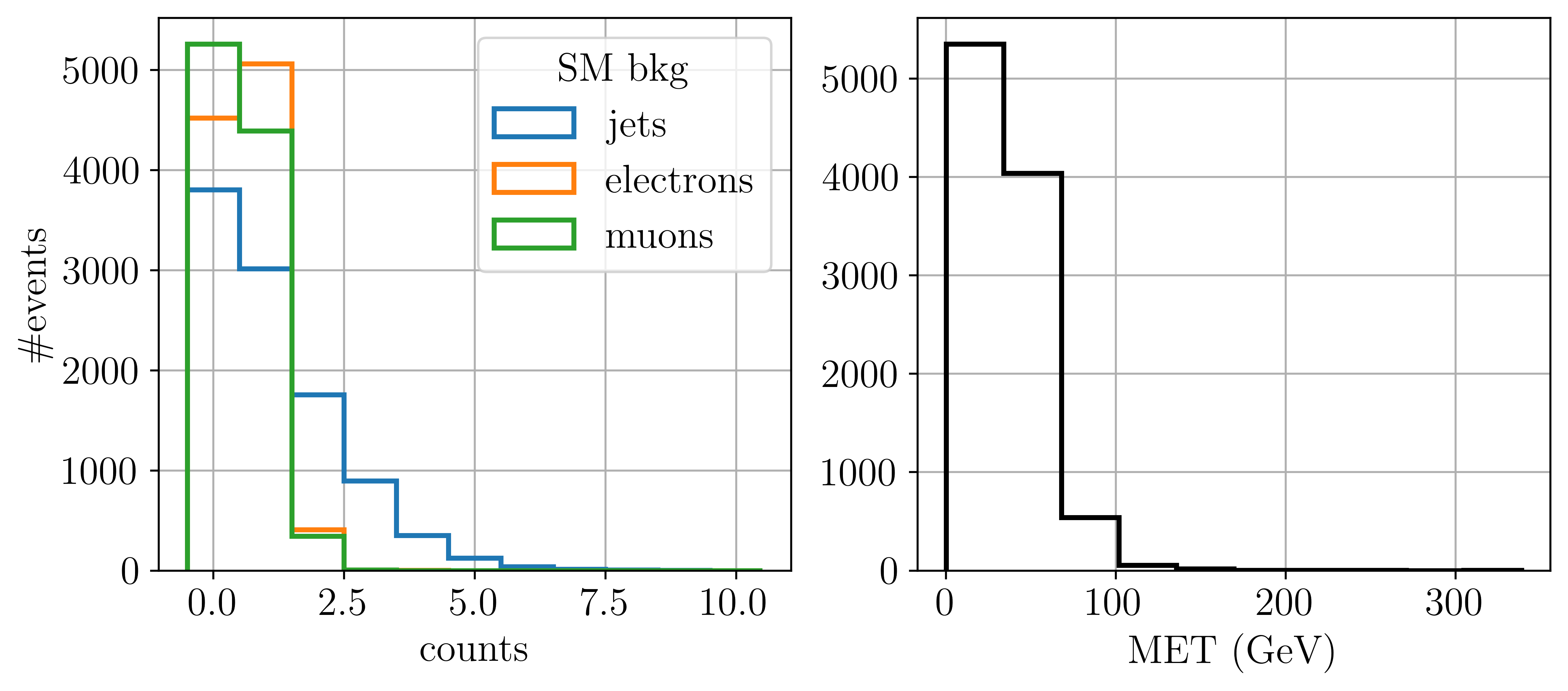}
    \caption{Histogram plots for the number of jets, electrons, and muons in the background SM dataset, and for the MET.}
    \label{fig:data_sm}
\end{figure}

The different processes contributing to the backgrounds are not labeled in the dataset, but they are:
\begin{itemize}
\itemsep0pt
    \item $59.2$\%: inclusive $W$-boson production with $W\!\rightarrow\! l\nu$, $l\!=\!e,\mu,\tau$
    \item $33.8$\%: QCD multijet production
    \item $6.7$\%: inclusive $Z$-boson production with $Z\!\rightarrow\!ll$
    \item $0.3$\%: $t\bar{t}$ production with at least one $t\!\rightarrow\!W^+b\!\rightarrow l\nu b$.
\end{itemize}
Plots showing the numbers of each particle type and the amount of MET in each event are shown in Fig.~\ref{fig:data_sm}.
We can see that events contain almost equal numbers of electrons and muons, due to the flavour universal decays of the $W$ and $Z$.
There are also many events with high-multiplicity jets arising from the QCD multijet background.

\subsection*{The signal datasets}
\begin{figure}[H]
    \includegraphics[width=\textwidth]{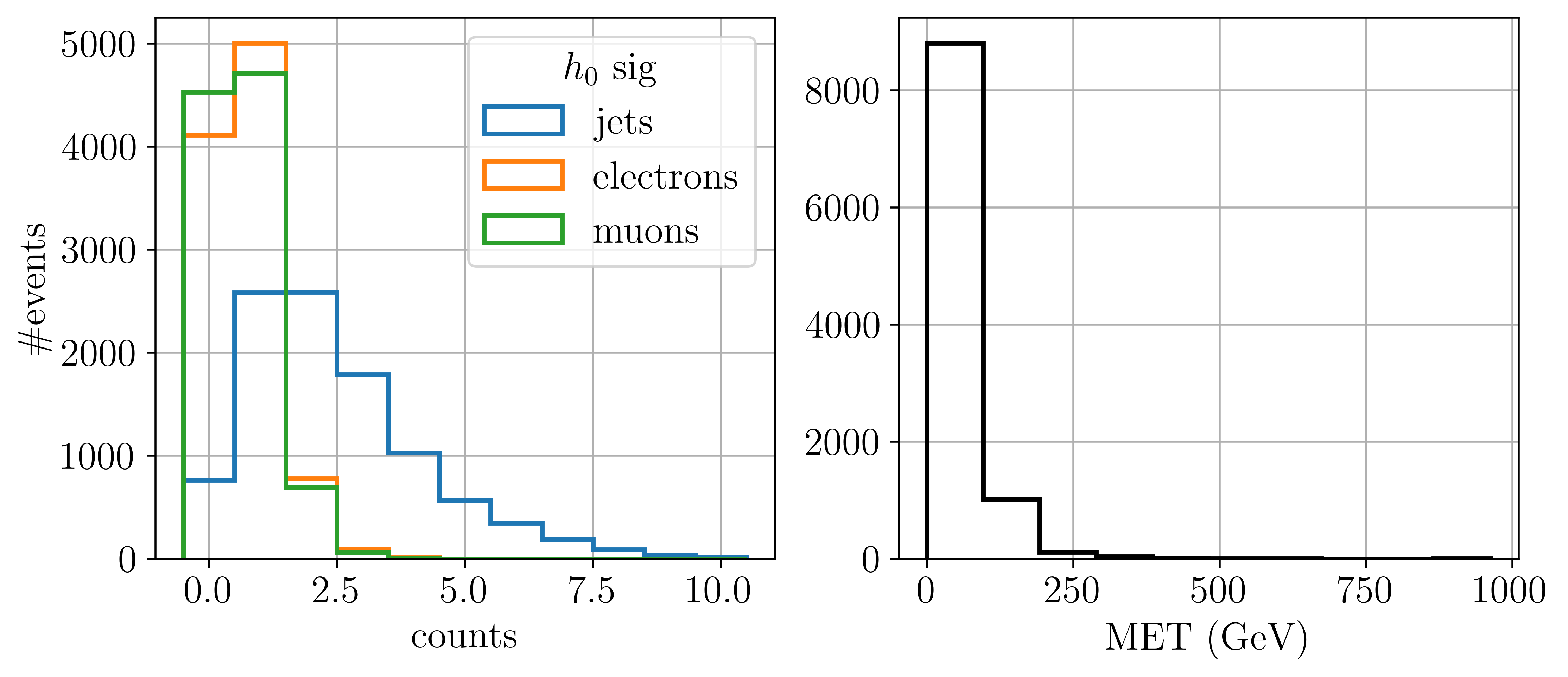}
    \caption{Histograms plots for the number of jets, electrons, and muons in the signal $h_0$ dataset, and for the MET.}
    \label{fig:data_h0}
\end{figure}

\begin{figure}[H]
    \includegraphics[width=\textwidth]{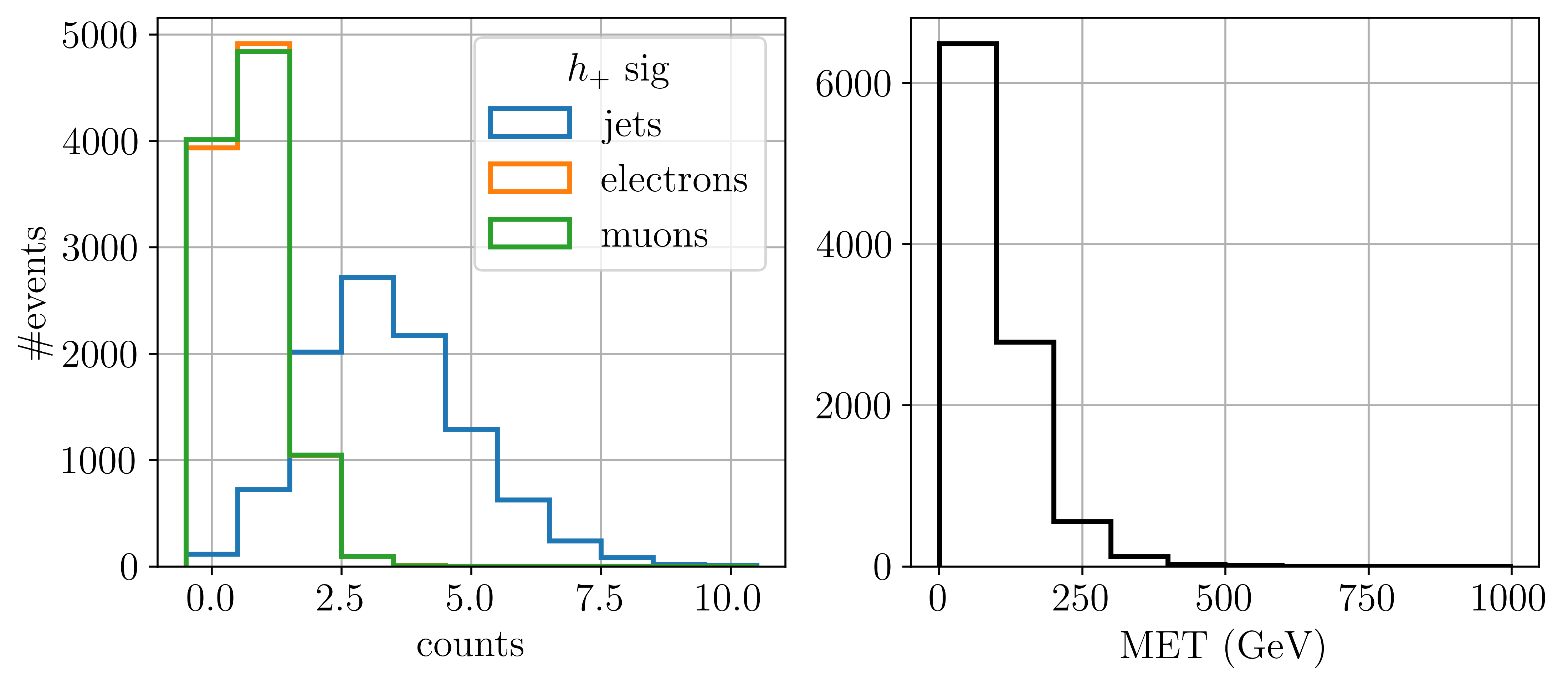}
    \caption{Histograms plots for the number of jets, electrons, and muons in the signal $h_+$ dataset, and for the MET.}
    \label{fig:data_hc}
\end{figure}

\begin{figure}[H]
    \includegraphics[width=\textwidth]{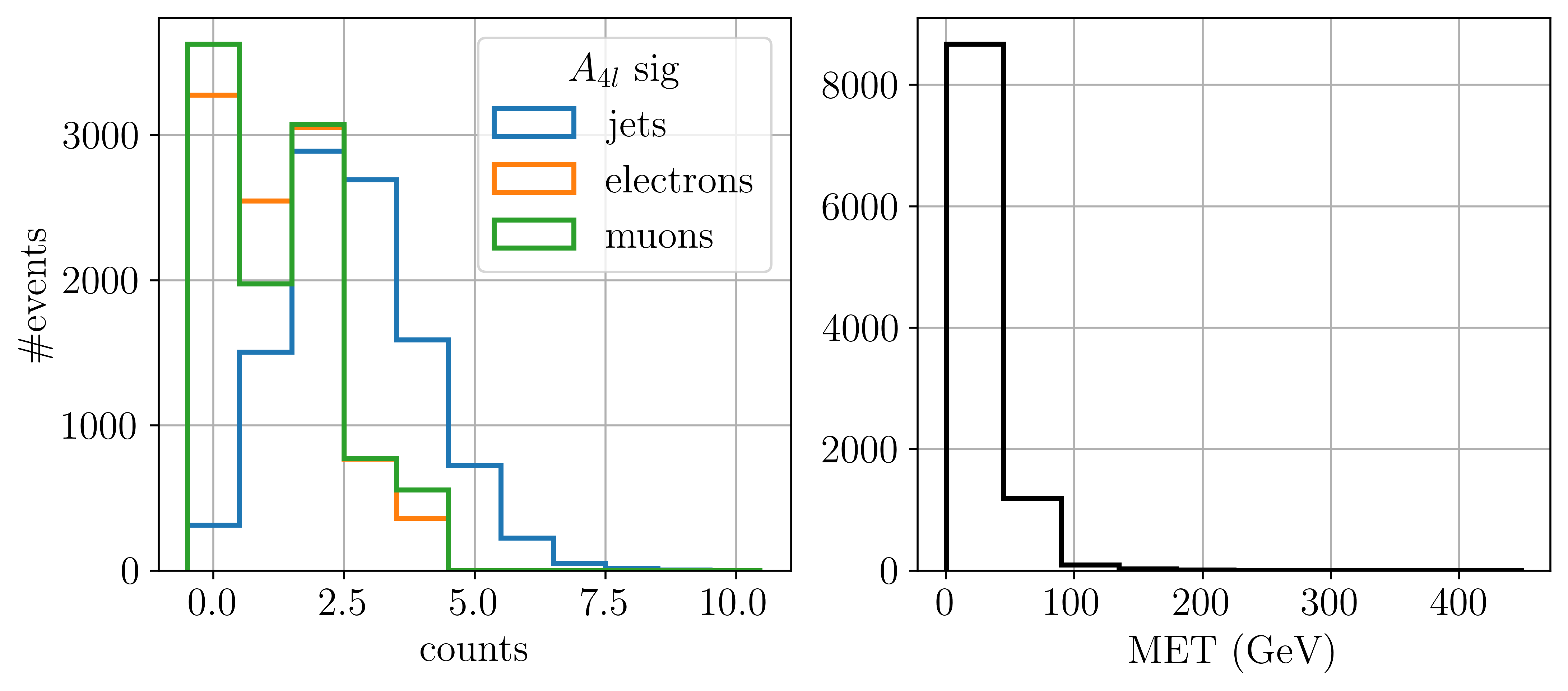}
    \caption{Histograms plots for the number of jets, electrons, and muons in the signal $A_{4l}$ dataset, and for the MET.}
    \label{fig:data_a4l}
\end{figure}

\begin{figure}[H]
    \includegraphics[width=\textwidth]{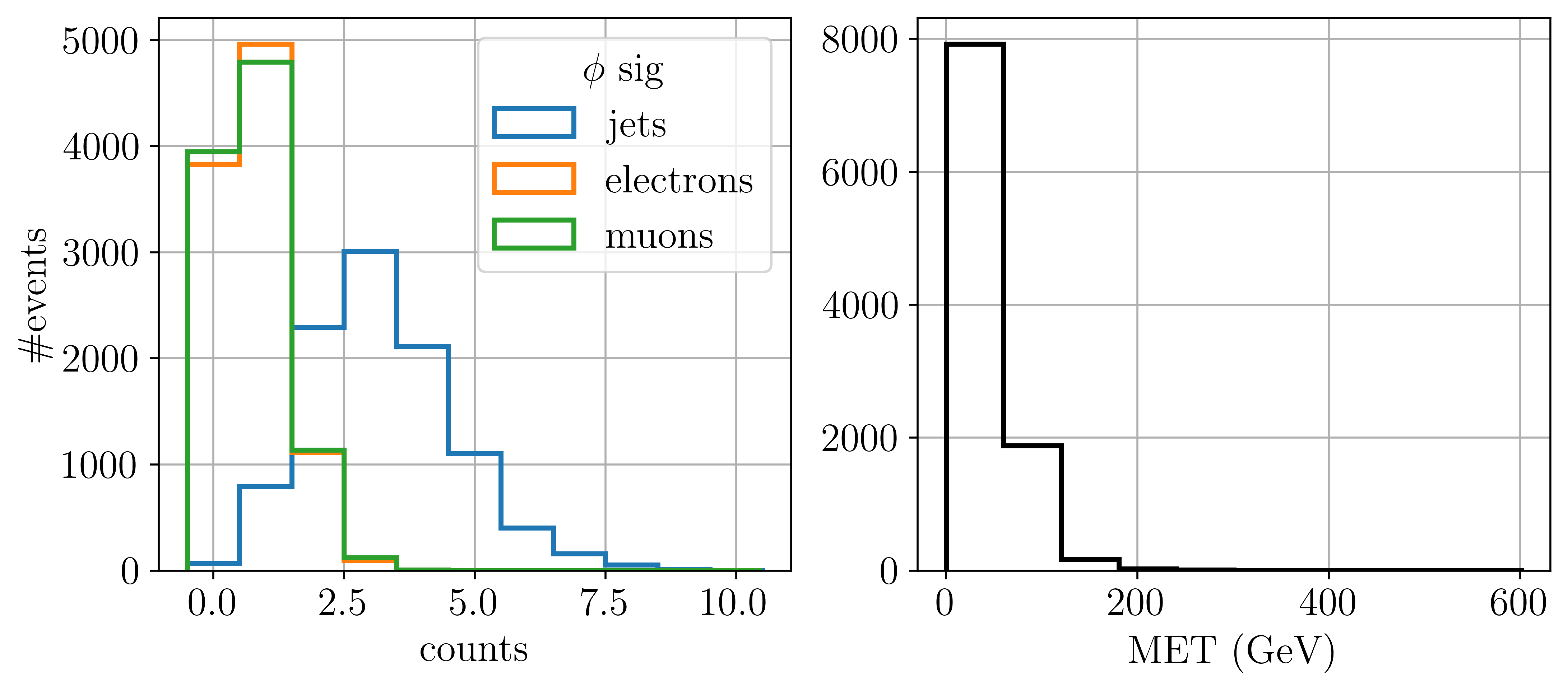}
    \caption{Histograms plots for the number of jets, electrons, and muons in the signal $\phi$ dataset, and for the MET.}
    \label{fig:data_lq}
\end{figure}

The signal dataset contains four different signal models with different underlying BSM physics and different final-state properties, they are:
\begin{itemize}
    \itemsep0pt
    \item a leptoquark (LQ, $\phi$): mass $80$ GeV, $\phi\!\rightarrow\! b\tau$
    \item a neutral scalar boson ($A$): mass $50$ GeV, $A\!\rightarrow\!Z^*Z^*\!\rightarrow\! llll$
    \item a scalar boson $h$: mass $60$ GeV, $h\!\rightarrow\!\tau\tau$
    \item a charged scalar boson $h^+$: mass $60$ GeV, $h^+\!\rightarrow\!\tau\nu$.
\end{itemize}
Plots showing the number of each particle type and the amount of MET in each event are illustrated in Fig.~\ref{fig:data_h0}, \ref{fig:data_hc}, \ref{fig:data_a4l}, and \ref{fig:data_lq}.
Compared to the SM background, all of the signals have higher multiplicity jet final states, while they generally have lower MET.
The signals also generally have more electrons and muons, especially the $A_{4l}$ signal.


\subsection*{Additional preprocessing}

We apply minimal preprocessing to the data: each of the $p_T$'s is linearly rescaled so that the maximum value of that feature in the background dataset equals $1.0$.
Additionally, the minimum value of each non-zero $p_T$ is shifted to $0.1$.
This means that if a third electron is present in an event, the minimum value of that entry in the input data will be $0.1$.
If there is no third electron in the event that entry will be $0.0$.
This preprocessing slightly improves performance because the presence of a particle in the final state, regardless of it's $p_T$, can be significant.

\section{Anomaly detection with SNNs}
\label{sec:snnaes}

The AutoEncoder architecture built with spiking neurons closely mirrors the conventional AutoEncoder. 
An input vector of length $M$ is mapped (encoded) to a latent space vector of length $D_z$ via an encoder SNN. 
This latent vector is then mapped (decoded) back to a vector of length $M$ through a decoder SNN.
The training objective is for the AutoEncoder to learn how to compress and reconstruct data it is trained on, such that outliers in the data are poorly reconstructed and can be identified.
Each event is represented by a vector of continuous floating-point numbers.
At each step in the forward pass, this vector is input to the input layer's neuron potentials.
After all $T$ steps, we read off the neuron potentials on the decoder output layer and treat this as the reconstructed input.
As already mentioned, we increase $u_{\text{thresh}}$ on the output layer to prevent those neurons from spiking.
The loss function that we use to measure the distance between the input and the reconstruction is the Mean Squared Error (MSE):
\begin{align}
L_T^{\text{AE}} &= \mathbb{E}_{x\sim p(x)}\left[ \left(x - f_T^{\text{AE}}\left(x\right)\right)^2 \right] \nonumber \\
&\simeq \sum_{i=1}^{N_{\text{batch}}} \left(x_i - f_T^{\text{AE}}\left(x_i\right)\right)^2
\end{align}
where the output of the complete encoder and decoder SNN architecture after the $T$ steps is represented by $f_T^{\text{AE}}\left(x\right)$.
The latent space of an SNN-AE differs fundamentally to that of a DNN-AE.
Due to the binary nature of the spikes, the information bottleneck is more constrained.
For a latent space of dimension $D_z$, there are exactly $2^{D_z}$ possible latent space configurations.
At each step the encoder SNN can produce different embeddings, so there are effectively $2^{T\cdot D_z}$ possible latent space configurations.

\begin{figure}[H]
    \centering
    \includegraphics[width=1.01\textwidth]{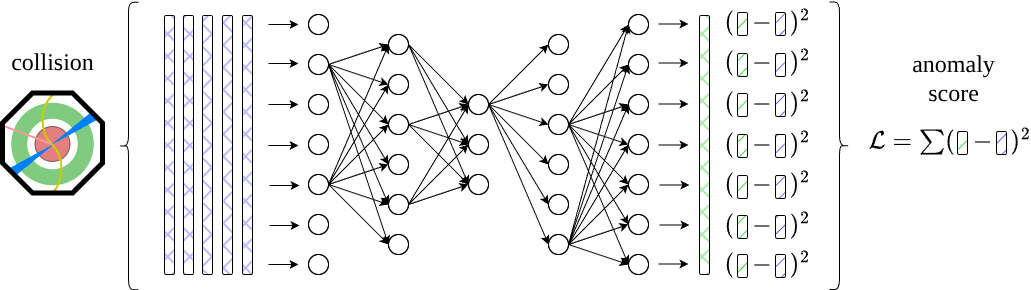}
    \caption{Schematic drawing of a forward-pass in the SNN AutoEncoder, indicating the sparse discrete firing between layers.}
\end{figure}

\subsection{Results}
\label{sec:results}

This work adopts relatively small AE architectures, considering the low-latency requirements of real-time applications. 
The default setup is an encoder with layers of $(19,24,12)$ neurons, a latent space with $4$ dimensions, and a decoder with layers of $(12,24,19)$ neurons. 
Performance is evaluated by comparing SNN-AEs against DNN-AEs with equivalent architectures. 
All models are implemented in \texttt{pytorch} \cite{Ansel_PyTorch_2_Faster_2024} with the SNN-specific functionality provided by \texttt{snntorch} \cite{eshraghian2021training}.
All networks are trained for $400$ epochs with the Adam optimizer \cite{Kingma:2014vow} with default parameters ($\beta_1\!=\!0.9$, $\beta_2\!=\!0.999$) and a learning rate of $0.001$.
Each model is trained on $100$k background events that are shuffled between each epoch.
The DNN-AE uses \texttt{ReLU} activations while the SNN-AE uses \texttt{Leaky} (leaky integrate and fire) activations implemented in \texttt{snntorch}.
Both the SNN-AE and DNN-AE are trained to minimize the MSE between the inputs and reconstructions.
There a few parameters unique to SNNs that we need to specify for the SNN-AE.
After some basic hyper-parameter searching we found that a good trade-off between computational efficiency and performance was to use $T\in[5,10]$, $u_{\text{thresh}}\!=\!1.2$ (with the exception of the final decoder layer), and a decay factor $\beta\!=\!0.9$.

\begin{figure}[H]
    \centering
    \begin{subfigure}[b]{0.45\textwidth}
        \centering
        \includegraphics[width=\textwidth]{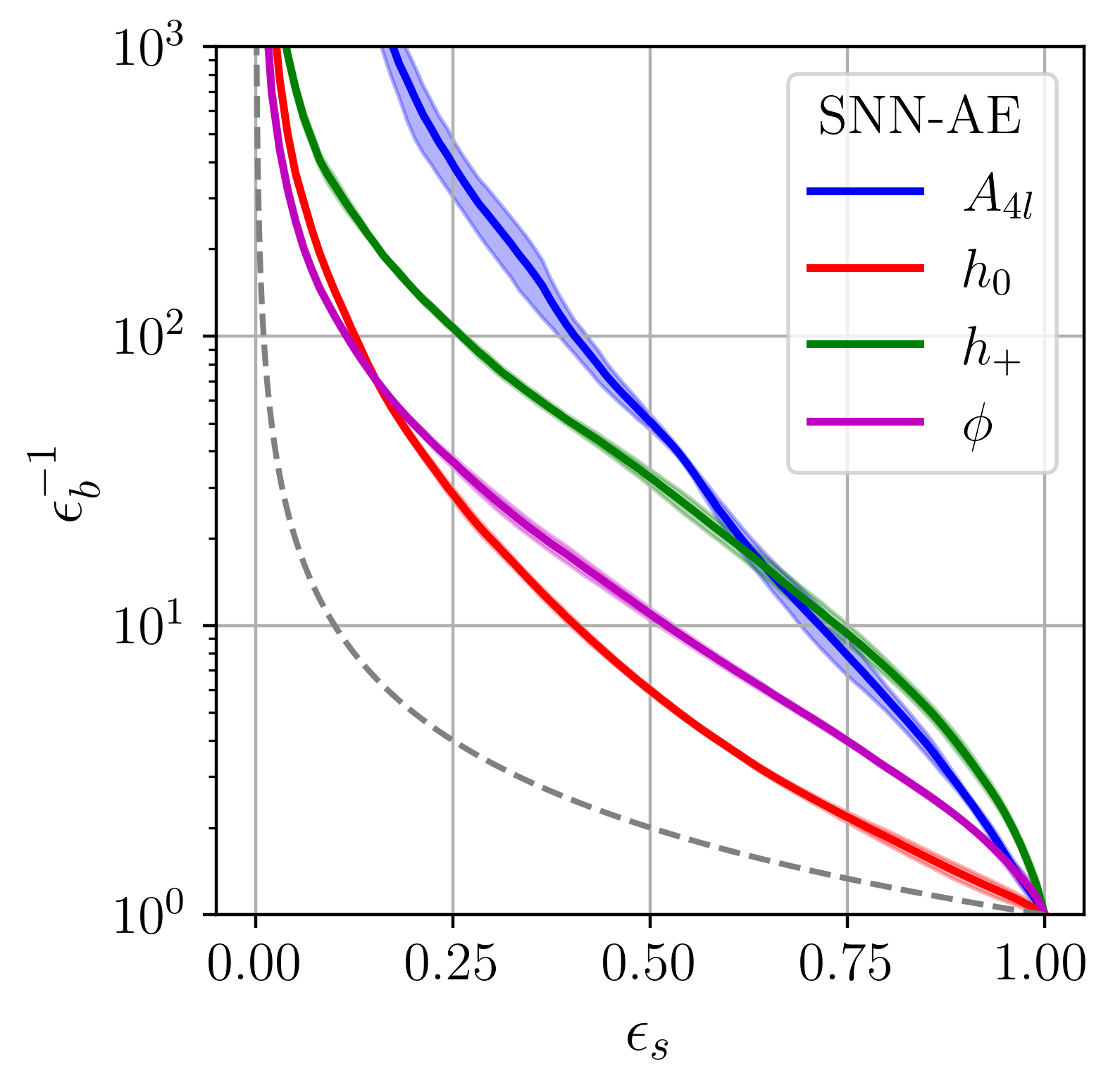}
    \end{subfigure}
    \begin{subfigure}[b]{0.45\textwidth}
        \centering
        \includegraphics[width=\textwidth]{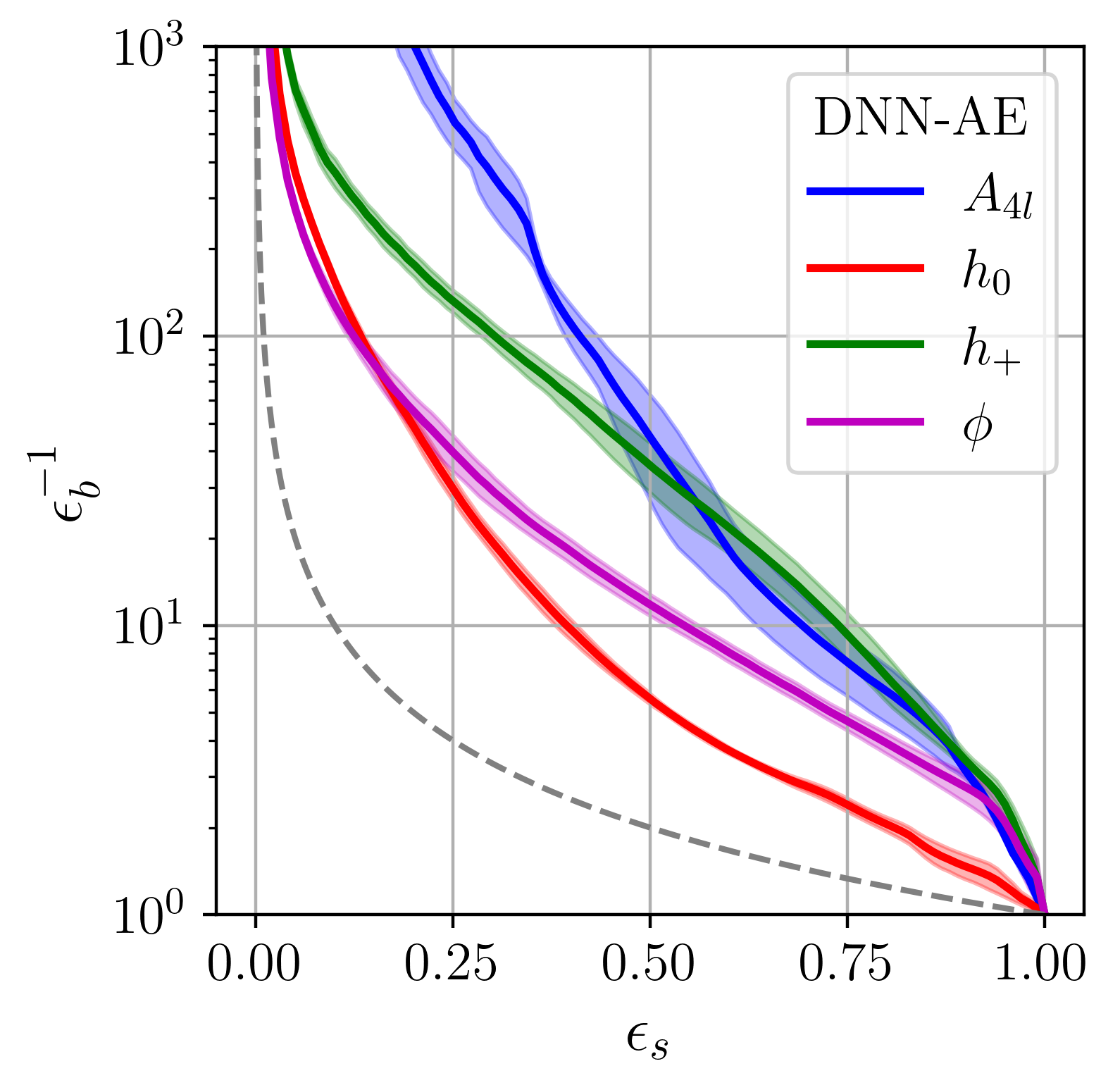}
    \end{subfigure}
    \caption{ROC curves comparing the SNN-AE ans DNN-AE performance on the CMS ADC signals.  All curves have been generated from the mean $\pm$ standard deviation over $5$ separate runs.
    \label{fig:results_roc1}}
\end{figure}
\begin{table}[H]
\centering
\begin{tabular}{l||l|l|l|l}
        & Loss  &   sig  & AUC & $\epsilon_b^{-1}(\epsilon_s\!=\!0.3)$ \\\hline\hline
SNN-AE  & $(20.24\pm0.35)\times10^{-5}$ & $h_0$  & $0.719\pm0.014$  &  $19.0\pm1.0$ \\
        &                            & $h_+$  & $0.899\pm0.006$    & $71.3\pm2.8$\\
        &                            & $A_{4l}$  & $0.880\pm0.006$    & $217.3\pm29.6$ \\
        &                            & $\phi$  & $0.802\pm0.011$    & $25.7\pm0.9$\\ \hline
DNN-AE  & $(5.19\pm0.95)\times10^{-5}$ & $h_0$   & $0.737\pm0.014$  & $19.4\pm1.6$ \\
        &                            & $h_+$  & $0.903\pm0.008$ & $101.8\pm9.6$ \\
        &                            & $A_{4l}$ & $0.890\pm0.011$ & $360.3\pm99.2$ \\
        &                            & $\phi$  & $0.852\pm0.014$ & $29.3\pm3.7$\\ \hline
\end{tabular}
\caption{Comparison of the SNN-AE to the DNN-AE performance on the CMS ADC.  All numbers have been generated from the mean $\pm$ standard deviation over $5$ separate runs.
\label{tab:results_tab1}}
\end{table}

\begin{figure}[H]
    \centering
    \begin{subfigure}[b]{0.45\textwidth}
        \centering
        \includegraphics[width=\textwidth]{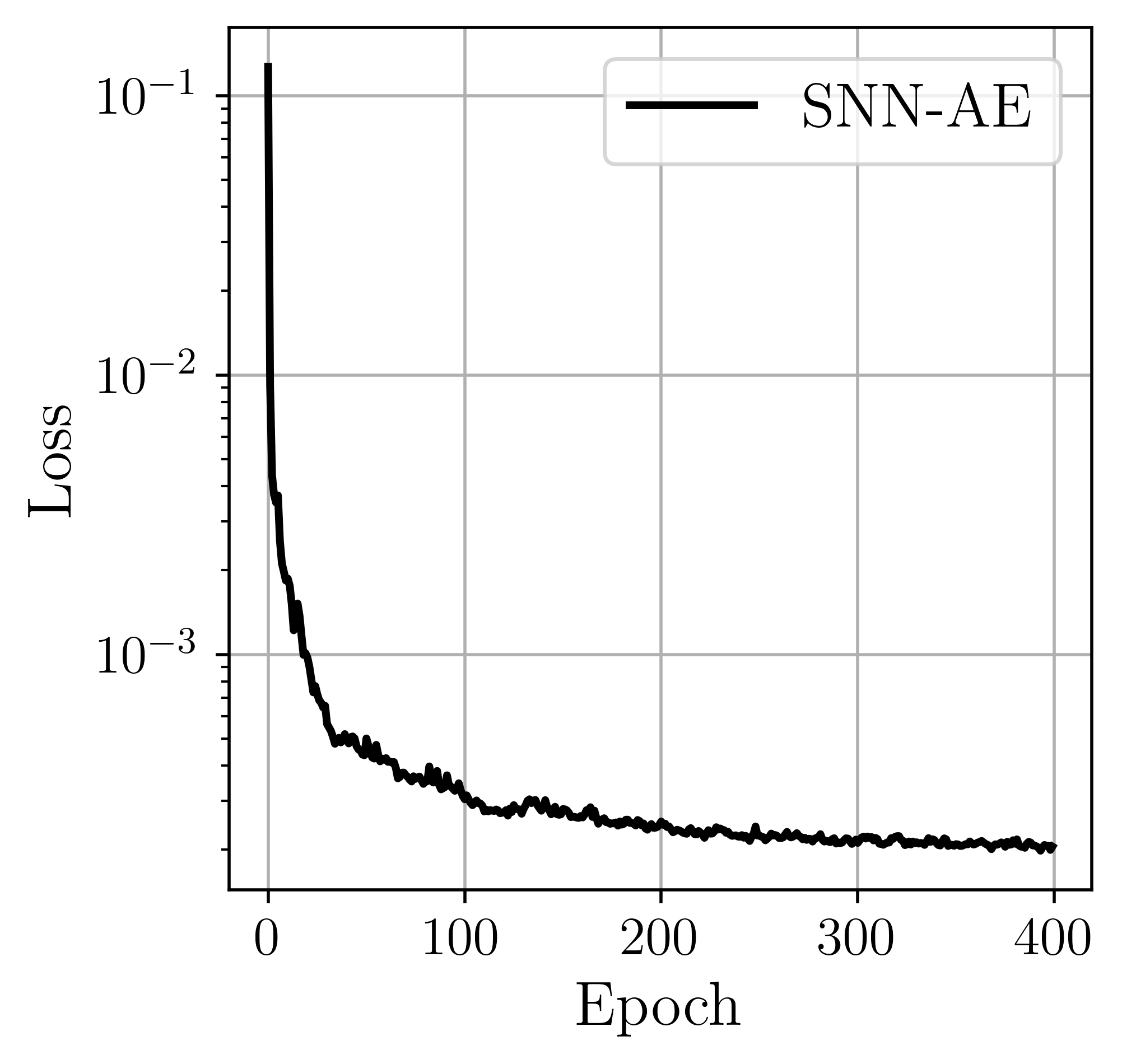}
    \end{subfigure}
    \begin{subfigure}[b]{0.45\textwidth}
        \centering
        \includegraphics[width=\textwidth]{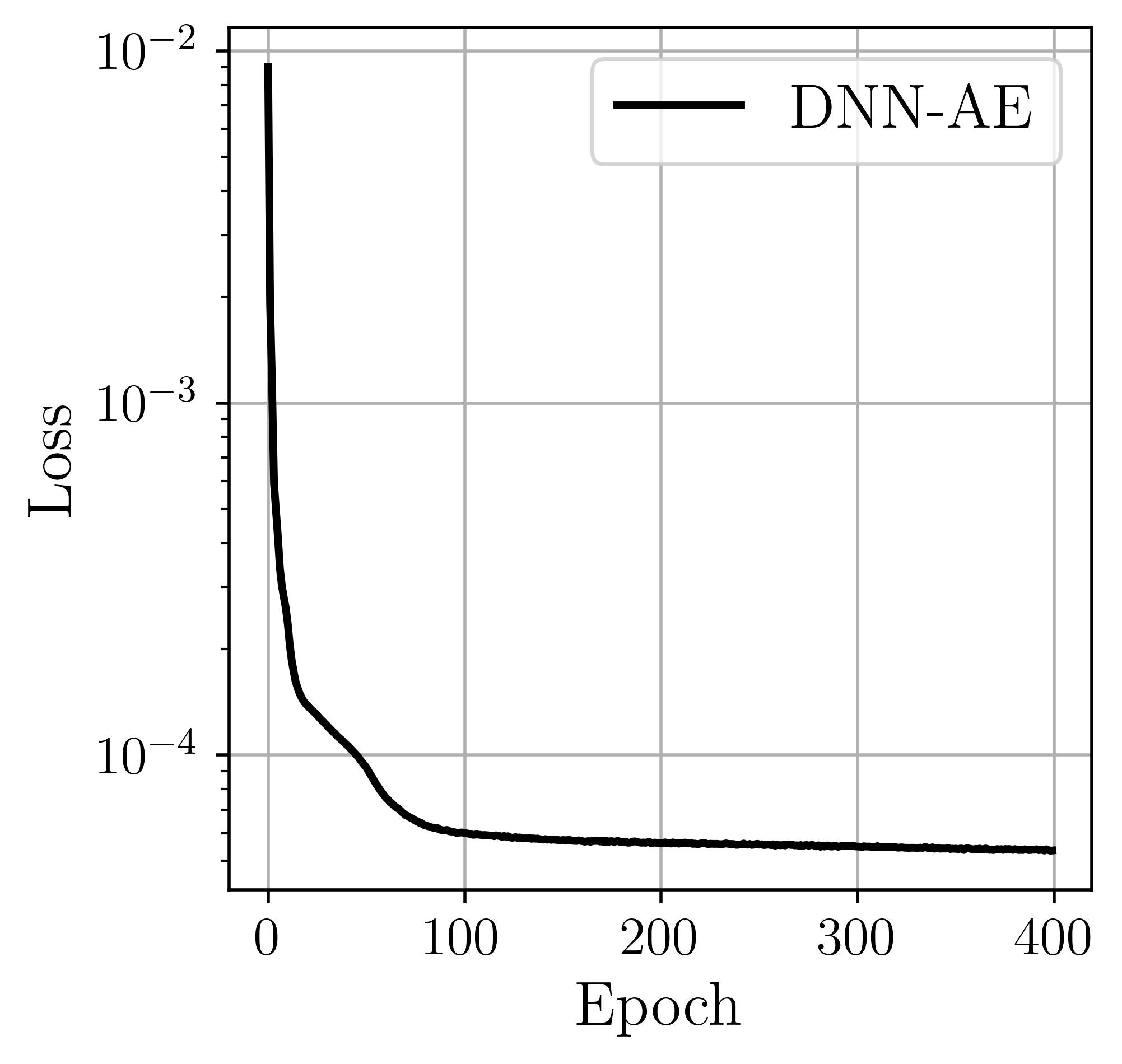}
    \end{subfigure}
    \caption{Comparison of the loss curves for two representative models of the trained SNN-AE to the DNN-AE results.
    \label{fig:loss_curves}}
\end{figure}

Fig.~\ref{fig:results_roc1} shows the ROC curve results on the CMS ADC data, with detailed performance metrics listed in Tab.~\ref{tab:results_tab1}.
As expected, the DNN-AE slightly outperforms the SNN-AE due to superior computational capacity. 
However the SNN-AE still performs very competitively, the differences are not significant.
In several cases, the differences between the SNN-AE and DNN-AE fall within the standard-deviation.
Moreover, the narrower error bands in the SNN-AE ROC curves suggest that they yield more consistent results across multiple runs.
Fig.~\ref{fig:loss_curves} compares the ROC curves for two representative models.
The DNN-AE does achieve a better loss value overall, and the SNN loss curve is noticeably noisier than the DNN-AE loss.
The cause of this may be that optimization techniques for SNNs are not yet as well developed than those for DNNs.

\subsection*{Latent space embeddings}

\begin{figure}[H]
    \centering
    \begin{subfigure}[b]{0.45\textwidth}
        \centering
        \includegraphics[width=\textwidth]{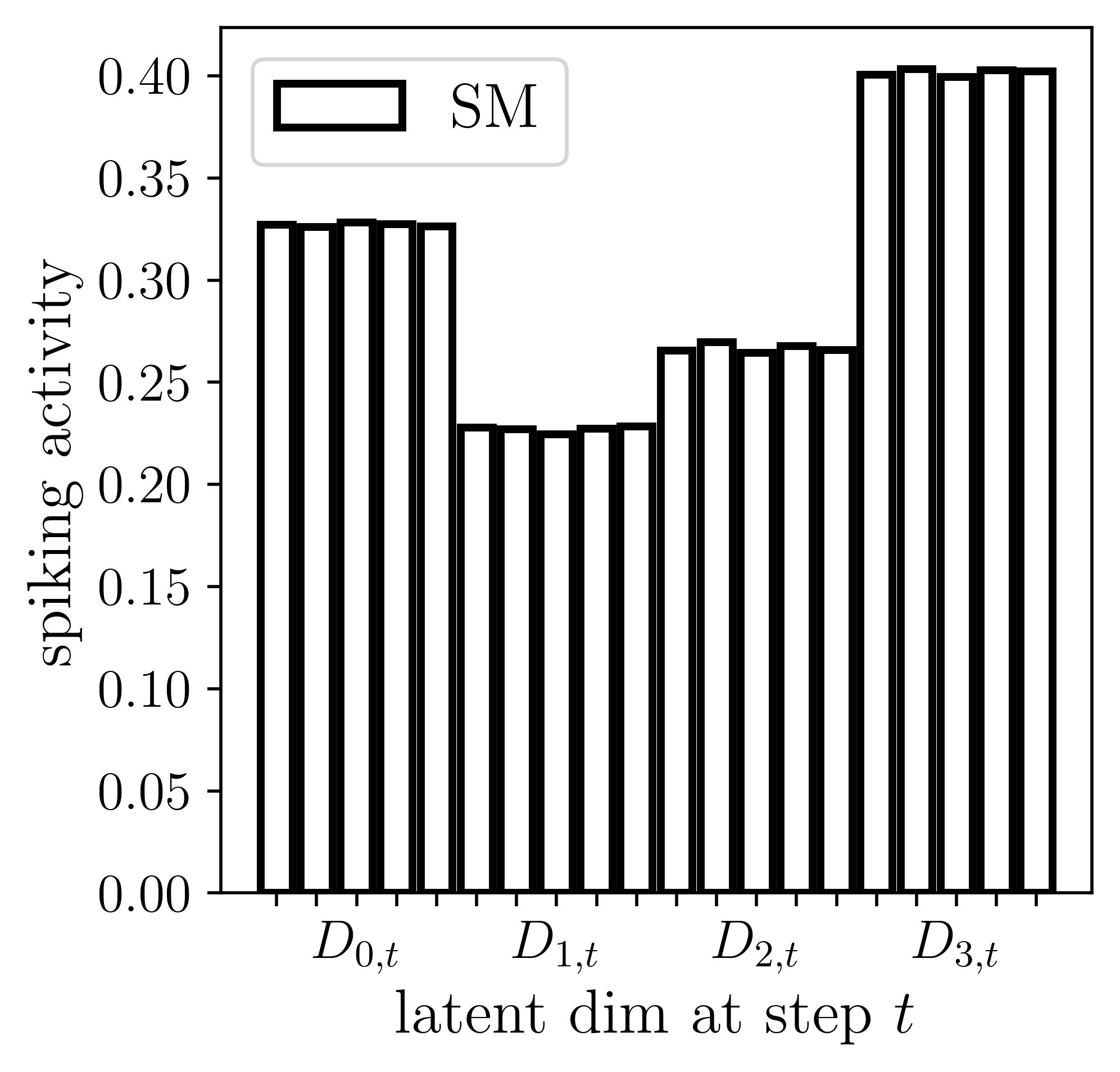}
    \end{subfigure}
    \caption{SNN-AE latent space embedding for the SM background events for $T\!=\!5$ steps.
    $D_{n,t}$ is the latent space representation of the $n^{\text{th}}$ dimension at step $t$.\label{fig:results_lse_sm}}
\end{figure}

\begin{figure}[H]
    \centering
    \begin{subfigure}[b]{0.45\textwidth}
        \centering
        \includegraphics[width=\textwidth]{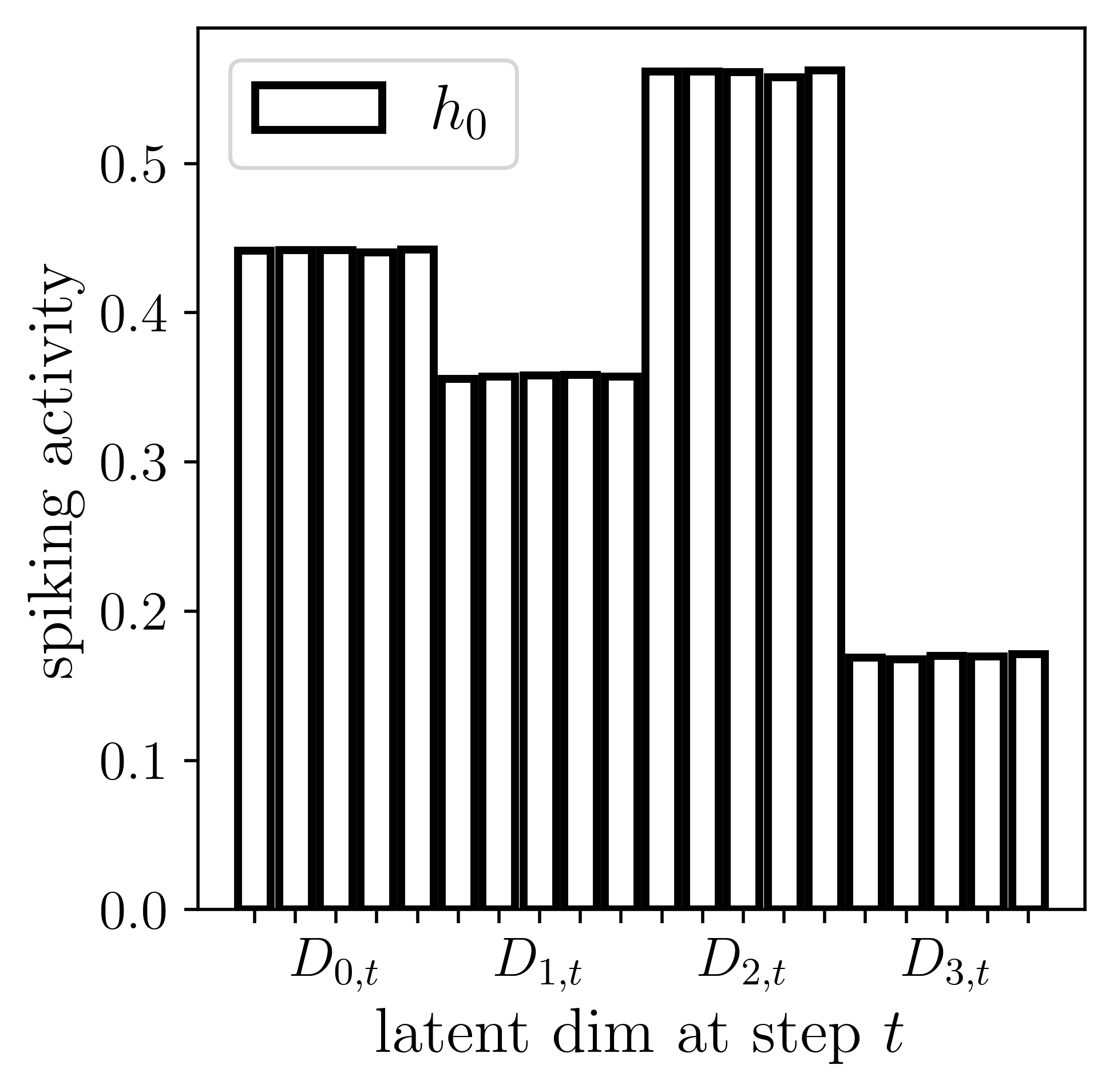}
    \end{subfigure}
    \begin{subfigure}[b]{0.45\textwidth}
        \centering
        \includegraphics[width=\textwidth]{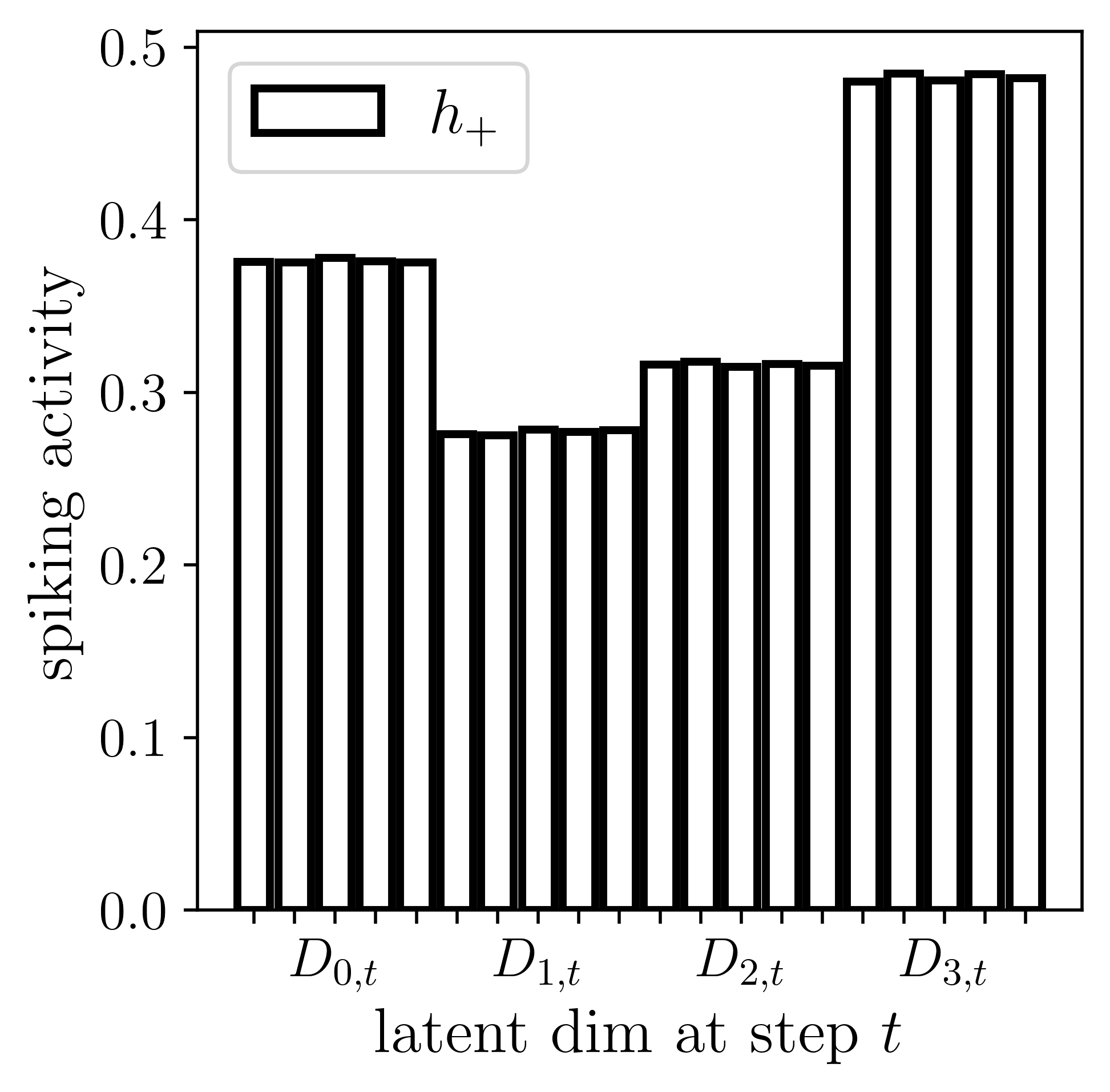}
    \end{subfigure}
    \begin{subfigure}[b]{0.45\textwidth}
        \centering
        \includegraphics[width=\textwidth]{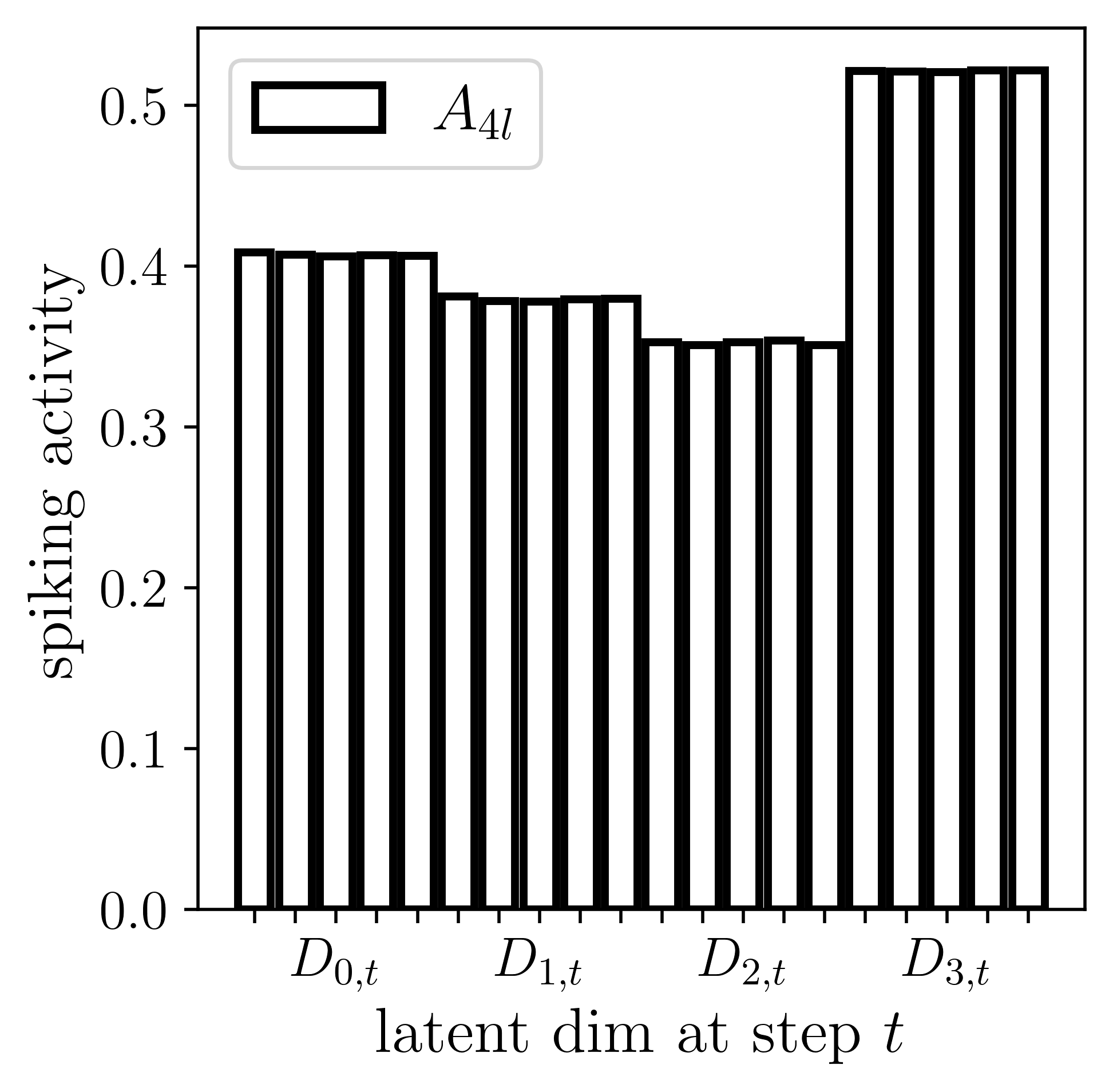}
    \end{subfigure}
    \begin{subfigure}[b]{0.45\textwidth}
        \centering
        \includegraphics[width=\textwidth]{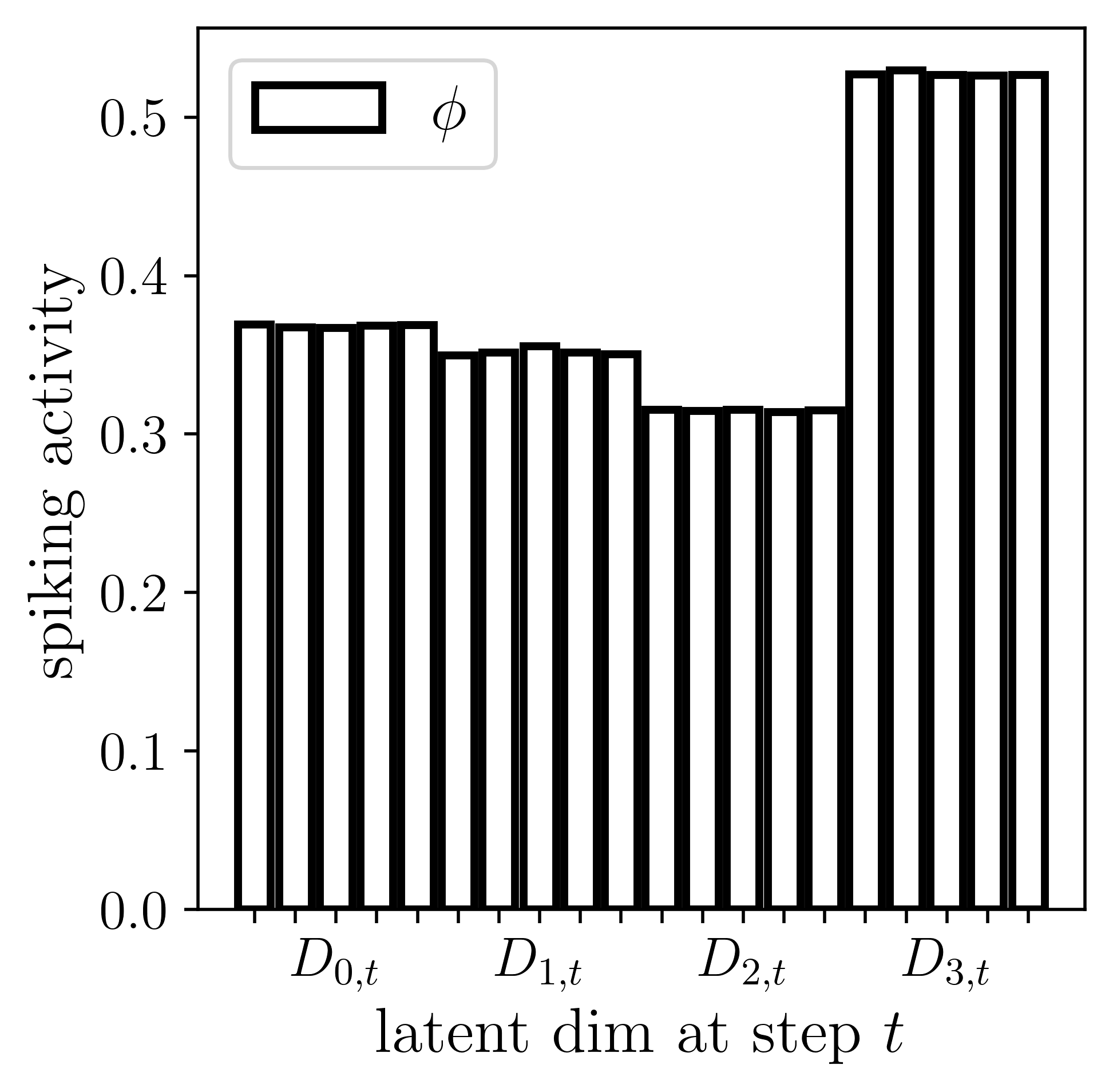}
    \end{subfigure}
    \caption{SNN-AE latent space embedding for the signal events for $T\!=\!5$ steps.
    $D_{n,t}$ is the latent space representation of the $n^{\text{th}}$ dimension at step $t$.\label{fig:results_lse_sig}}
\end{figure}

Studying latent space representations in AutoEncoders is a common way to understand how signal and background events are processed differently by the networks.
This is particularly interesting for the SNN-AE, since the latent space representations differ fundamentally from those of regular DNNs.
For the DNN-AE latent space representation for an event is a vector of floating-point numbers, while for the SNN-AE it is a vector of zeros (no spike) and ones (spike).
There are $T$ of these vectors for each event, each corresponding to one step in the forward pass, and they are not independent of each other.
Fig.~\ref{fig:results_lse_sm} and Fig.~\ref{fig:results_lse_sig} show the latent space embedding for the SM background events and signal events, respectively, for one of the trained models in Fig.~\ref{fig:results_roc1}.
There are $T\!=\!5$ steps in the forward-pass, therefore each latent dimension has $5$ entries in the plots. 
As expected, the latent representations for each event type are different.
But we also see that the representation for any particular event type is approximately the same for each step.
This is because we are averaging over the whole dataset, while in general, different events may spike in the same dimension at different steps.

\subsection*{Performance vs \#steps}

\begin{figure}
    \begin{subfigure}[b]{0.5\textwidth}
    \centering
    \includegraphics[width=1.0\linewidth]{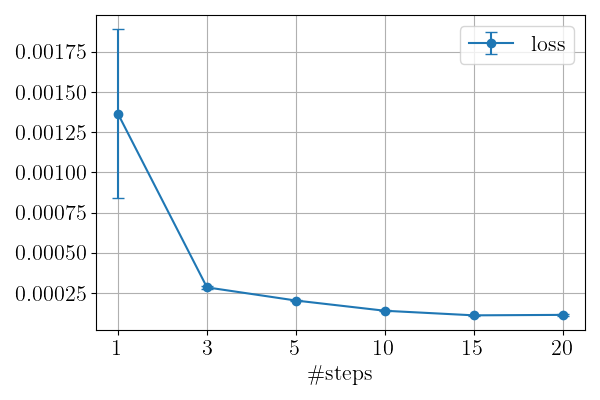}
    \end{subfigure}
    \begin{subfigure}[b]{0.5\textwidth}
    \centering
    \includegraphics[width=1.0\linewidth]{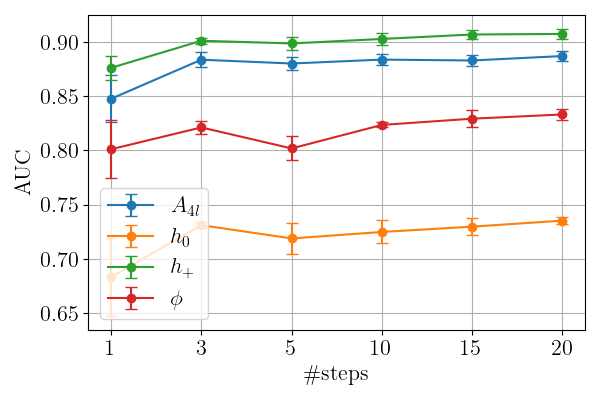}
    \end{subfigure}
    \caption{Here we show the final loss and AUC change in SNN-AEs where the number of steps in the forward-pass ($T$) is varied from $1$ to $20$.  All errors are calculated from the average over five models.\label{fig:vary-steps}}
\end{figure}

The SNNs process information internally in the form of discrete spikes. 
For an SNN with just one step, $T\!=\!1$, then the neuron potentials on the output layer get just one discrete update and the SNN-AE has a limited ability to reconstruct the input data.
As $T$ increases, the network's ability to make precise reconstructions grows, and so better loss values and better anomaly detection performance are expected. 
This exact pattern is shown in Fig.~\ref{fig:vary-steps} using the same SNN-AE architecture as before.
That is, an encoder with layers $(19,24,12)$, a decoder with layers $(12,24,19)$, and a latent space dimension of $4$.
The neural networks are trained using six different choices of $T$, $(1,3,5,10,15,20)$.
For each choice we train five different networks and take the mean and standard deviation to compare the results.
We notice that not only does the performance increase with more steps $T$, but the variance in the results between different runs generally decreases.
While more steps produces better results, the trade-off in the end is between performance and computational efficiency.

\subsection*{Performance vs latent dimension}

\begin{figure}[H]
    \begin{subfigure}[b]{0.5\textwidth}
    \centering
    \includegraphics[width=1.0\linewidth]{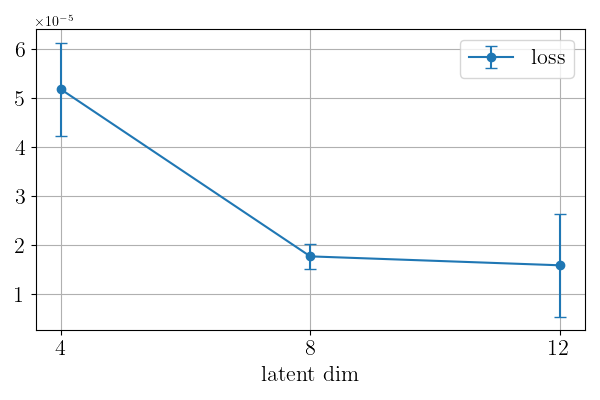}
    \end{subfigure}
    \begin{subfigure}[b]{0.5\textwidth}
    \centering
    \includegraphics[width=1.0\linewidth]{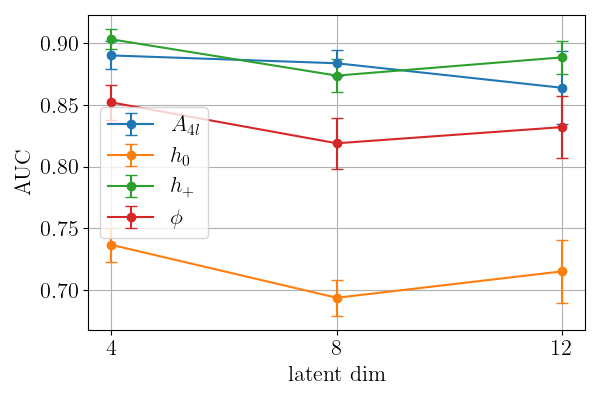}
    \end{subfigure}
    \caption{Here we show the final loss and AUC change in DNN-AEs where the latent space dimension is varied.  All errors are calculated from the average over five models.\label{fig:dnn-vary-latent}}
\end{figure}

\begin{figure}[H]
    \begin{subfigure}[b]{0.5\textwidth}
    \centering
    \includegraphics[width=1.0\linewidth]{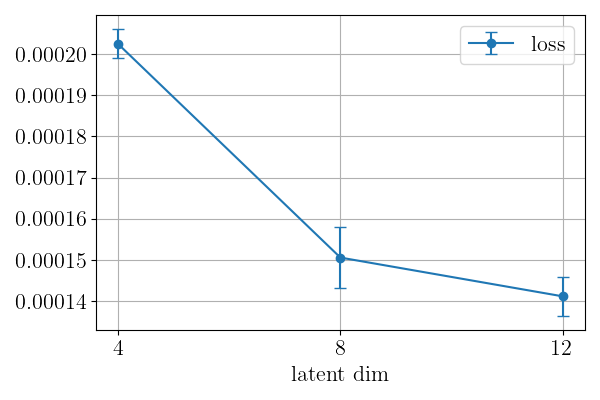}
    \end{subfigure}
    \begin{subfigure}[b]{0.5\textwidth}
    \centering
    \includegraphics[width=1.0\linewidth]{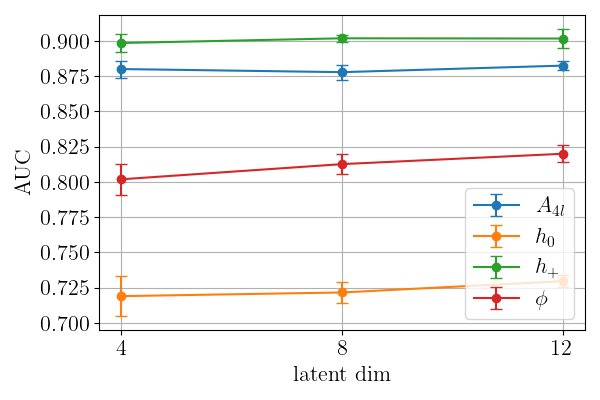}
    \end{subfigure}
    \caption{Here we show the final loss and AUC change in SNN-AEs where the latent space dimension is varied.  All errors are calculated from the average over five models.\label{fig:snn-vary-latent}}
\end{figure}

Next, we examine how performance varies with the size of the latent space in both the SNN-AE and the DNN-AE. 
Choosing an appropriate latent dimension is non-trivial.
On one hand, a larger latent space enables the network to obtain better reconstructions and a better overall loss.
While on the other hand, a latent space that is too large can lead to `outlier reconstruction'.
Where despite not bein trained on anomalous events, the network can still partially reconstruct them. 
This hinders anomaly detection performance.
For this analysis the encoder and decoder layers remain the same, and the number of steps is kept at $T\!=\!5$.
We compare latent space dimensions of $(4,8,12)$.
For each choice of latent space, five different models are trained, and the mean and standard deviation of the performance metrics are calculated for comparison. 
Fig.~\ref{fig:dnn-vary-latent} and Fig.~\ref{fig:snn-vary-latent} show the results.
As expected, the losses decrease as the size of the latent space increases. 
However anomaly detection performance for the DNN-AE generally worsens with larger latent dimension and the variance grows. 
The SNN-AE behaves better as the latent dimension increases. 
The average AUC remains stable or even increases, and the variance is much smaller than with the DNN-AE. 

\subsection*{Performance with limited data}

\begin{figure}[H]
    \begin{subfigure}[b]{0.5\textwidth}
    \centering
    \includegraphics[width=1.0\linewidth]{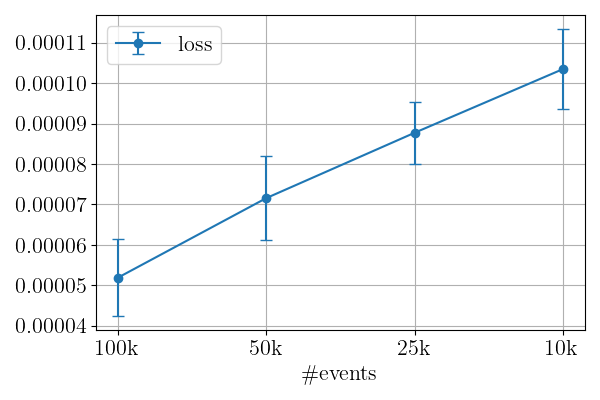}
    \end{subfigure}
    \begin{subfigure}[b]{0.5\textwidth}
    \centering
    \includegraphics[width=1.0\linewidth]{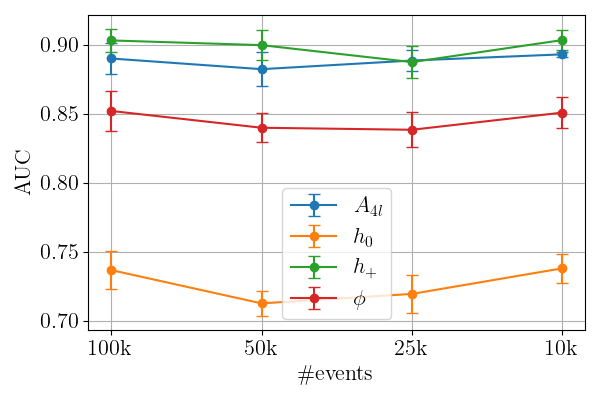}
    \end{subfigure}
    \caption{Here we show the final loss and AUC change in DNN-AEs where the number of events we train on is varied.  All errors are calculated from the average over five models.\label{fig:dnn-vary-num-events}}
\end{figure}

\begin{figure}[H]
    \begin{subfigure}[b]{0.5\textwidth}
    \centering
    \includegraphics[width=1.0\linewidth]{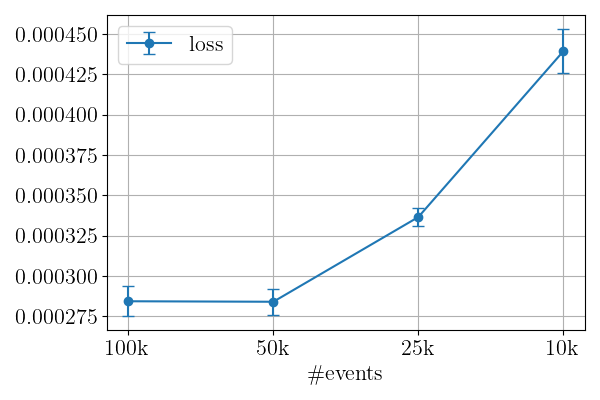}
    \end{subfigure}
    \begin{subfigure}[b]{0.5\textwidth}
    \centering
    \includegraphics[width=1.0\linewidth]{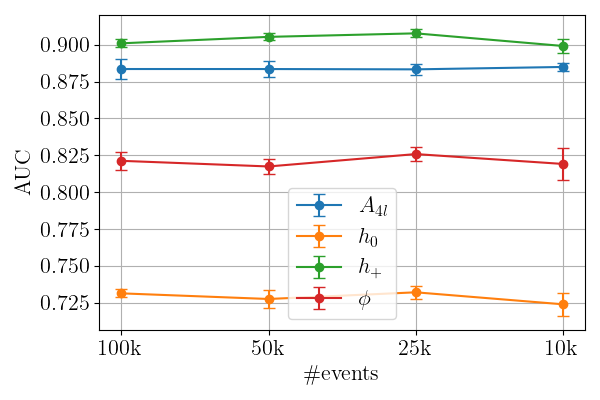}
    \end{subfigure}
    \caption{Here we show the final loss and AUC change in SNN-AEs where the number of events we train on is varied.  All errors are calculated from the average over five models.\label{fig:snn-vary-num-events}}
\end{figure}

Finally we assess how the networks perform when trained on limited data.
Until now we have used $100$k background events for training. 
Three additional scenarios are considered with ($50$k, $25$k, $10$k) training events.
To ensure the same number of updates to the network weights, the models are trained for ($800$, $1600$, $4000$) epochs, respectively.
The same architecture with $T\!=\!5$ is used, and again five models are trained for each case.
Fig.~\ref{fig:dnn-vary-num-events} and Fig.~\ref{fig:snn-vary-num-events} show the results.
As expected, limited training data leads to an increase in the overall loss, despite the number of weight updates remaining constant.
However compared to the DNN-AE, the SNN-AE loss appears more robust to modest decreases in the training data size.
While the loss increases with limited data, anomaly detection performance across all four signal models appears relatively stable. 
The variance in the AUC for the SNN-AE models remains smaller than the variance for the DNN-AE models in all cases except with $10$k events, where the variances are approximately equal.

\section{Conclusions}
\label{sec:conclusions}
This work introduces a Spiking Neural Network AutoEncoder (SNN-AE) architecture for anomaly detection at the LHC\footnote{Code for this project will be maintained at \url{https://github.com/bmdillon/spike-hep}.}. 
We presented an overview of SNNs and their integration with the AutoEncoder architecture and evaluated the SNN-AE performance on the CMS ADC dataset.
Despite their limited computational complexity, the SNN-AE performed similarly to the DNN-AE on all benchmarks.
In particular, the SNN-AEs demonstrated greater robustness to variations in network initialization, latent space size, and training dataset size.

There are several promising directions for future research.
One involves deploying the SNN-AE to dedicated hardware, where more realistic performance and efficiency tests can be performed. 
Hardware choices include both FPGA and neuromorphic hardware, such as Intel Loihi chips. 
Both deployments will require more in-depth work on network optimization and on how the physics data is represented. 
Another direction involves applying the SNN-AE, or SNNs in general, to other physics or particle physics scenarios. 
The use of deep learning in the physical sciences is relatively young and has until now been mostly reserved for offline analysis of data. 
In recent years, impressive advances in FastML and neuromorphic hardware have opened new opportunities for online deep learning algorithms in experiments, most notably at the CMS experiment. 
It is not unrealistic to expect future experiments to be designed with specific FastML hardware in mind, pushing the limits of what can be achieved.  
Now is the time to explore and refine the potential applications of these technologies. 

\subsection*{Acknowledgements}

We are grateful for use of the computing resources from the Northern Ireland High Performance Computing (NI-HPC) service funded by EPSRC (EP/T022175).

\printbibliography

@article{Mikuni:2021nwn,
    author = "Mikuni, Vinicius and Nachman, Benjamin and Shih, David",
    title = "{Online-compatible unsupervised nonresonant anomaly detection}",
    eprint = "2111.06417",
    archivePrefix = "arXiv",
    primaryClass = "cs.LG",
    doi = "10.1103/PhysRevD.105.055006",
    journal = "Phys. Rev. D",
    volume = "105",
    number = "5",
    pages = "055006",
    year = "2022"
}

@inproceedings{Duarte:2022hdp,
    author = "Duarte, Javier and Tran, Nhan and Hawks, Ben and Herwig, Christian and Muhizi, Jules and Prakash, Shvetank and Reddi, Vijay Janapa",
    title = "{FastML Science Benchmarks: Accelerating Real-Time Scientific Edge Machine Learning}",
    booktitle = "{5th Conference on Machine Learning and Systems}",
    eprint = "2207.07958",
    archivePrefix = "arXiv",
    primaryClass = "cs.LG",
    reportNumber = "FERMILAB-CONF-22-534-PPD-SCD",
    month = "7",
    year = "2022"
}

@article{Kosters:2022amb,
    author = {K\"osters, Dominique J. and others},
    title = "{Benchmarking energy consumption and latency for neuromorphic computing in condensed matter and particle physics}",
    eprint = "2209.10481",
    archivePrefix = "arXiv",
    primaryClass = "cs.ET",
    doi = "10.1063/5.0116699",
    journal = "APL Mach. Learn.",
    volume = "1",
    number = "1",
    pages = "016101",
    year = "2023"
}

@article{Jia:2024ysq,
    author = "Jia, Haoyi and Dave, Abhilasha and Gonski, Julia and Herbst, Ryan",
    title = "{Analysis of Hardware Synthesis Strategies for Machine Learning in Collider Trigger and Data Acquisition}",
    eprint = "2411.11678",
    archivePrefix = "arXiv",
    primaryClass = "physics.ins-det",
    month = "11",
    year = "2024"
}

@Article{Dillon:2021nxw,
  author        = {Dillon, Barry M. and Plehn, Tilman and Sauer, Christof and Sorrenson, Peter},
  journal       = {SciPost Phys.},
  title         = {{Better Latent Spaces for Better Autoencoders}},
  year          = {2021},
  pages         = {061},
  volume        = {11},
  archiveprefix = {arXiv},
  doi           = {10.21468/SciPostPhys.11.3.061},
  eprint        = {2104.08291},
  primaryclass  = {hep-ph},
}

@article{Buss_2023,
   title={What’s anomalous in LHC jets?},
   volume={15},
   ISSN={2542-4653},
   url={http://dx.doi.org/10.21468/SciPostPhys.15.4.168},
   DOI={10.21468/scipostphys.15.4.168},
   number={4},
   journal={SciPost Physics},
   publisher={Stichting SciPost},
   author={Buss, Thorsten and Dillon, Barry M. and Finke, Thorben and Krämer, Michael and Morandini, Alessandro and Mück, Alexander and Oleksiyuk, Ivan and Plehn, Tilman},
   year={2023},
   month=oct }

@article{Dillon:2022mkq,
    author = {Dillon, Barry M. and Favaro, Luigi and Plehn, Tilman and Sorrenson, Peter and Kr\"amer, Michael},
    title = "{A normalized autoencoder for LHC triggers}",
    eprint = "2206.14225",
    archivePrefix = "arXiv",
    primaryClass = "hep-ph",
    doi = "10.21468/SciPostPhysCore.6.4.074",
    journal = "SciPost Phys. Core",
    volume = "6",
    pages = "074",
    year = "2023"
}

@Article{Dillon:2020quc,
  author        = {Dillon, B. M. and Faroughy, D. A. and Kamenik, J. F. and Szewc, M.},
  journal       = {JHEP},
  title         = {{Learning the latent structure of collider events}},
  year          = {2020},
  pages         = {206},
  volume        = {10},
  archiveprefix = {arXiv},
  doi           = {10.1007/JHEP10(2020)206},
  eprint        = {2005.12319},
  primaryclass  = {hep-ph},
}

@Article{Farina:2018fyg,
  author        = {Farina, Marco and Nakai, Yuichiro and Shih, David},
  journal       = {Phys. Rev. D},
  title         = {{Searching for New Physics with Deep Autoencoders}},
  year          = {2020},
  pages         = {075021},
  volume        = {101},
  archiveprefix = {arXiv},
  doi           = {10.1103/PhysRevD.101.075021},
  eprint        = {1808.08992},
  primaryclass  = {hep-ph},
}

@article{Khosa_2023,
   title={Anomaly Awareness},
   volume={15},
   ISSN={2542-4653},
   url={http://dx.doi.org/10.21468/SciPostPhys.15.2.053},
   DOI={10.21468/scipostphys.15.2.053},
   number={2},
   journal={SciPost Physics},
   publisher={Stichting SciPost},
   author={Khosa, Charanjit Kaur and Sanz, Veronica},
   year={2023},
   month=aug }

@article{Banda:2025nrv,
    author = "Banda, Adam and Khosa, Charanjit K. and Sanz, Veronica",
    title = "{Strengthening Anomaly Awareness}",
    eprint = "2504.11520",
    archivePrefix = "arXiv",
    primaryClass = "hep-ph",
    month = "4",
    year = "2025"
}

@article{Cheng_2023,
   title={Variational autoencoders for anomalous jet tagging},
   volume={107},
   ISSN={2470-0029},
   url={http://dx.doi.org/10.1103/PhysRevD.107.016002},
   DOI={10.1103/physrevd.107.016002},
   number={1},
   journal={Physical Review D},
   publisher={American Physical Society (APS)},
   author={Cheng, Taoli and Arguin, Jean-François and Leissner-Martin, Julien and Pilette, Jacinthe and Golling, Tobias},
   year={2023},
   month=jan }

@article{Govorkova:2021utb,
    author = "Govorkova, Ekaterina and others",
    title = "{Autoencoders on field-programmable gate arrays for real-time, unsupervised new physics detection at 40 MHz at the Large Hadron Collider}",
    eprint = "2108.03986",
    archivePrefix = "arXiv",
    primaryClass = "physics.ins-det",
    reportNumber = "FERMILAB-PUB-21-487-CMS, FERMILAB-PUB-21-487-CMS",
    doi = "10.1038/s42256-022-00441-3",
    journal = "Nature Mach. Intell.",
    volume = "4",
    pages = "154--161",
    year = "2022"
}

@inproceedings{Schuman:2017kdp,
    author = "Schuman, Catherine D. and Potok, Thomas E. and Young, Steven and Patton, Robert and Perdue, Gabriel and Chakma, Gangotree and Wyer, Austin and Rose, Garrett S.",
    title = "{Neuromorphic Computing for Temporal Scientific Data Classification}",
    reportNumber = "FERMILAB-CONF-17-659-CD",
    year = "2017"
}

@article{Aliaga_2014,
   title={Design, calibration, and performance of the MINERvA detector},
   volume={743},
   ISSN={0168-9002},
   url={http://dx.doi.org/10.1016/j.nima.2013.12.053},
   DOI={10.1016/j.nima.2013.12.053},
   journal={Nuclear Instruments and Methods in Physics Research Section A: Accelerators, Spectrometers, Detectors and Associated Equipment},
   publisher={Elsevier BV},
   author={MINERvA Collaboration},
   year={2014},
   month=apr, pages={130–159} }

@article{Kulkarni:2023lpb,
    author = "Kulkarni, Shruti R. and others",
    title = "{On-Sensor Data Filtering using Neuromorphic Computing for High Energy Physics Experiments}",
    eprint = "2307.11242",
    archivePrefix = "arXiv",
    primaryClass = "cs.NE",
    reportNumber = "FERMILAB-PUB-23-431-PPD",
    month = "7",
    year = "2023"
}

@article{Coradin:2025ees,
    author = "Coradin, Emanuele and Cufino, Fabio and Awais, Muhammad and Dorigo, Tommaso and Lupi, Enrico and Porcu, Eleonora and Raj, Jinu and Sandin, Fredrik and Tosi, Mia",
    collaboration = "MODE",
    title = "{Unsupervised Particle Tracking with Neuromorphic Computing}",
    eprint = "2502.06771",
    archivePrefix = "arXiv",
    primaryClass = "hep-ex",
    doi = "10.3390/particles8020040",
    month = "2",
    year = "2025"
}

@article{2021:cmsmusic,
   title={MUSiC: a model-unspecific search for new physics in proton–proton collisions at $\sqrt{s} = 13\,\text {TeV}$},
   volume={81},
   ISSN={1434-6052},
   url={http://dx.doi.org/10.1140/epjc/s10052-021-09236-z},
   DOI={10.1140/epjc/s10052-021-09236-z},
   number={7},
   journal={The European Physical Journal C},
   publisher={Springer Science and Business Media LLC},
   author={CMS Collaboration},
   year={2021},
   month=jul }

@article{Aaboud:2018ufy,
      author         = "Aaboud, Morad and others",
      title          = "{A strategy for a general search for new phenomena using data-derived signal regions and its application within the ATLAS experiment}",
      collaboration  = "ATLAS",
      journal        = "Eur. Phys. J.",
      volume         = "C79",
      year           = "2019",
      pages          = "120",
      doi            = "10.1140/epjc/s10052-019-6540-y",
      eprint         = "1807.07447",
      archivePrefix  = "arXiv",
      primaryClass   = "hep-ex",
      reportNumber   = "CERN-EP-2018-070",
      SLACcitation   = "%%CITATION = ARXIV:1807.07447;%%"
}

@Article{Heimel:2018mkt,
  author        = {Heimel, Theo and Kasieczka, Gregor and Plehn, Tilman and Thompson, Jennifer M.},
  journal       = {SciPost Phys.},
  title         = {{QCD or What?}},
  year          = {2019},
  pages         = {030},
  volume        = {6},
  archiveprefix = {arXiv},
  doi           = {10.21468/SciPostPhys.6.3.030},
  eprint        = {1808.08979},
  primaryclass  = {hep-ph},
}

@article{Roy:2019jae,
      author         = "Roy, Tuhin S. and Vijay, Aravind H.",
      title          = "{A robust anomaly finder based on autoencoder}",
      year           = "2019",
      eprint         = "1903.02032",
      archivePrefix  = "arXiv",
      primaryClass   = "hep-ph",
      reportNumber   = "TIFR/TH/19-4",
      SLACcitation   = "%%CITATION = ARXIV:1903.02032;%%"
}

@Article{Nachman:2020lpy,
  author        = {Nachman, Benjamin and Shih, David},
  journal       = {Phys. Rev. D},
  title         = {{Anomaly Detection with Density Estimation}},
  year          = {2020},
  pages         = {075042},
  volume        = {101},
  archiveprefix = {arXiv},
  doi           = {10.1103/PhysRevD.101.075042},
  eprint        = {2001.04990},
  primaryclass  = {hep-ph},
}

@article{Hallin_2022,
	doi = {10.1103/physrevd.106.055006},
	url = {https://doi.org/10.1103%2Fphysrevd.106.055006},
	year = 2022,
	month = {sep},
	publisher = {American Physical Society ({APS})},
	volume = {106},
	number = {5},
	author = {Anna Hallin and Joshua Isaacson and Gregor Kasieczka and Claudius Krause and Benjamin Nachman and Tobias Quadfasel and Matthias Schlaffer and David Shih and Manuel Sommerhalder},
	title = {Classifying anomalies through outer density estimation},
	journal = {Physical Review D}
}

@article{Cerri:2018anq,
      author         = "Cerri, Olmo and Nguyen, Thong Q. and Pierini, Maurizio
                        and Spiropulu, Maria and Vlimant, Jean-Roch",
      title          = "{Variational Autoencoders for New Physics Mining at the Large Hadron Collider}",
      journal        = "JHEP",
      volume         = "05",
      year           = "2019",
      pages          = "036",
      doi            = "10.1007/JHEP05(2019)036",
      eprint         = "1811.10276",
      archivePrefix  = "arXiv",
      primaryClass   = "hep-ex",
      SLACcitation   = "%%CITATION = ARXIV:1811.10276;%%"
}

@article{Blance:2019ibf,
      author         = "Blance, Andrew and Spannowsky, Michael and Waite, Philip",
      title          = "{Adversarially-trained autoencoders for robust unsupervised new physics searches}",
      journal        = "JHEP",
      volume         = "10",
      year           = "2019",
      pages          = "047",
      doi            = "10.1007/JHEP10(2019)047",
      eprint         = "1905.10384",
      archivePrefix  = "arXiv",
      primaryClass   = "hep-ph",
      reportNumber   = "IPPP/19/41",
      SLACcitation   = "%%CITATION = ARXIV:1905.10384;%%"
}

@article{Hajer:2018kqm,
      author         = "Hajer, Jan and Li, Ying-Ying and Liu, Tao and Wang, He",
      title          = "{Novelty Detection Meets Collider Physics}",
      year           = "2018",
      eprint         = "1807.10261",
      archivePrefix  = "arXiv",
      primaryClass   = "hep-ph",
      SLACcitation   = "%%CITATION = ARXIV:1807.10261;%%"
}

@article{Collins:2018epr,
      author         = "Collins, Jack H. and Howe, Kiel and Nachman, Benjamin",
      title          = "{Anomaly Detection for Resonant New Physics with Machine Learning}",
      journal        = "Phys. Rev. Lett.",
      volume         = "121",
      year           = "2018",
      number         = "24",
      pages          = "241803",
      doi            = "10.1103/PhysRevLett.121.241803",
      eprint         = "1805.02664",
      archivePrefix  = "arXiv",
      primaryClass   = "hep-ph",
      reportNumber   = "FERMILAB-PUB-18-180-T",
      SLACcitation   = "%%CITATION = ARXIV:1805.02664;%%"
}

@article{Collins:2019jip,
      author         = "Collins, Jack H. and Howe, Kiel and Nachman, Benjamin",
      title          = "{Extending the search for new resonances with machine learning}",
      journal        = "Phys. Rev.",
      volume         = "D99",
      year           = "2019",
      number         = "1",
      pages          = "014038",
      doi            = "10.1103/PhysRevD.99.014038",
      eprint         = "1902.02634",
      archivePrefix  = "arXiv",
      primaryClass   = "hep-ph",
      reportNumber   = "FERMILAB-PUB-18-733-T",
      SLACcitation   = "%%CITATION = ARXIV:1902.02634;%%"
}

@Article{Dillon:2019cqt,
  author        = {Dillon, Barry M. and Faroughy, Darius A. and Kamenik, Jernej F.},
  journal       = {Phys. Rev. D},
  title         = {{Uncovering latent jet substructure}},
  year          = {2019},
  pages         = {056002},
  volume        = {100},
  archiveprefix = {arXiv},
  doi           = {10.1103/PhysRevD.100.056002},
  eprint        = {1904.04200},
  primaryclass  = {hep-ph},
}

@article{Duarte:2018ite,
    author = "Duarte, Javier and others",
    archivePrefix = "arXiv",
    doi = "10.1088/1748-0221/13/07/P07027",
    eprint = "1804.06913",
    journal = "JINST",
    number = "07",
    pages = "P07027",
    primaryClass = "physics.ins-det",
    reportNumber = "FERMILAB-PUB-18-089-E",
    title = "{Fast inference of deep neural networks in FPGAs for particle physics}",
    volume = "13",
    year = "2018"
}

@Article{Metodiev:2017vrx,
  author        = {Metodiev, Eric M. and Nachman, Benjamin and Thaler, Jesse},
  journal       = {JHEP},
  title         = {{Classification without labels: Learning from mixed samples in high energy physics}},
  year          = {2017},
  pages         = {174},
  volume        = {10},
  archiveprefix = {arXiv},
  doi           = {10.1007/JHEP10(2017)174},
  eprint        = {1708.02949},
  primaryclass  = {hep-ph},
}

@article{Kingma:2014vow,
    author = "Kingma, Diederik P. and Ba, Jimmy",
    title = "{Adam: A Method for Stochastic Optimization}",
    eprint = "1412.6980",
    archivePrefix = "arXiv",
    primaryClass = "cs.LG",
    month = "12",
    year = "2014"
}

@article{Kasieczka:2021xcg,
    author = "Kasieczka, Gregor and others",
    title = "{The LHC Olympics 2020: A Community Challenge for Anomaly Detection in High Energy Physics}",
    eprint = "2101.08320",
    archivePrefix = "arXiv",
    primaryClass = "hep-ph",
    month = "1",
    year = "2021"
}

@article{Ostdiek_2022,
   title={Deep Set Auto Encoders for Anomaly Detection in Particle Physics},
   volume={12},
   ISSN={2542-4653},
   url={http://dx.doi.org/10.21468/SciPostPhys.12.1.045},
   DOI={10.21468/scipostphys.12.1.045},
   number={1},
   journal={SciPost Physics},
   publisher={Stichting SciPost},
   author={Ostdiek, Bryan},
   year={2022},
   month=jan }

@article{Govorkova:2021hqu,
    author = "Govorkova, Ekaterina and Puljak, Ema and Aarrestad, Thea and Pierini, Maurizio and Wo\'zniak, Kinga Anna and Ngadiuba, Jennifer",
    title = "{LHC physics dataset for unsupervised New Physics detection at 40 MHz}",
    eprint = "2107.02157",
    archivePrefix = "arXiv",
    primaryClass = "physics.data-an",
    reportNumber = "FERMILAB-PUB-21-338-CMS",
    doi = "10.1038/s41597-022-01187-8",
    journal = "Sci.  Data",
    volume = "9",
    pages = "118",
    year = "2022"
}

@article{Faucett:2022zie,
    author = "Faucett, Taylor and Hsu, Shih-Chieh and Whiteson, Daniel",
    title = "{Learning to identify semi-visible jets}",
    eprint = "2208.10062",
    archivePrefix = "arXiv",
    primaryClass = "hep-ph",
    doi = "10.1007/JHEP12(2022)132",
    journal = "JHEP",
    volume = "12",
    pages = "132",
    year = "2022"
}

@software{fastml_hls4ml,
  author       = {{FastML Team}},
  title        = {fastmachinelearning/hls4ml},
  year         = 2025,
  publisher    = {Zenodo},
  version      = {v1.1.0},
  doi          = {10.5281/zenodo.1201549},
  url          = {https://github.com/fastmachinelearning/hls4ml}
}

@article{Krause:2025qnl,
    author = "Krause, Claudius and Wang, Daohan and Winterhalder, Ramon",
    title = "{BitHEP -- The Limits of Low-Precision ML in HEP}",
    eprint = "2504.03387",
    archivePrefix = "arXiv",
    primaryClass = "hep-ph",
    reportNumber = "HEPHY-ML-25-02",
    month = "4",
    year = "2025"
}

@article{Dillon:2021gag,
    author = "Dillon, Barry M. and Kasieczka, Gregor and Olischlager, Hans and Plehn, Tilman and Sorrenson, Peter and Vogel, Lorenz",
    title = "{Symmetries, safety, and self-supervision}",
    eprint = "2108.04253",
    archivePrefix = "arXiv",
    primaryClass = "hep-ph",
    doi = "10.21468/SciPostPhys.12.6.188",
    journal = "SciPost Phys.",
    volume = "12",
    number = "6",
    pages = "188",
    year = "2022"
}

@article{Dillon:2023zac,
    author = "Dillon, Barry M. and Favaro, Luigi and Feiden, Friedrich and Modak, Tanmoy and Plehn, Tilman",
    title = "{Anomalies, representations, and self-supervision}",
    eprint = "2301.04660",
    archivePrefix = "arXiv",
    primaryClass = "hep-ph",
    doi = "10.21468/SciPostPhysCore.7.3.056",
    journal = "SciPost Phys. Core",
    volume = "7",
    pages = "056",
    year = "2024"
}

@article{Matos:2024ggs,
    author = "Matos, Gabriel and Busch, Elena and Park, Ki Ryeong and Gonski, Julia",
    title = "{Semi-supervised permutation invariant particle-level anomaly detection}",
    eprint = "2408.17409",
    archivePrefix = "arXiv",
    primaryClass = "hep-ph",
    doi = "10.1007/JHEP05(2025)116",
    journal = "JHEP",
    volume = "05",
    pages = "116",
    year = "2025"
}

@article{Favaro:2023xdl,
    author = {Favaro, Luigi and Kr\"amer, Michael and Modak, Tanmoy and Plehn, Tilman and R\"uschkamp, Jan},
    title = "{Semi-visible jets, energy-based models, and self-supervision}",
    eprint = "2312.03067",
    archivePrefix = "arXiv",
    primaryClass = "hep-ph",
    doi = "10.21468/SciPostPhys.18.2.042",
    journal = "SciPost Phys.",
    volume = "18",
    number = "2",
    pages = "042",
    year = "2025"
}

@article{Dillon:2022tmm,
    author = "Dillon, Barry M. and Mastandrea, Radha and Nachman, Benjamin",
    title = "{Self-supervised anomaly detection for new physics}",
    eprint = "2205.10380",
    archivePrefix = "arXiv",
    primaryClass = "hep-ph",
    doi = "10.1103/PhysRevD.106.056005",
    journal = "Phys. Rev. D",
    volume = "106",
    number = "5",
    pages = "056005",
    year = "2022"
}

@article{Raine:2022hht,
    author = "Raine, John Andrew and Klein, Samuel and Sengupta, Debajyoti and Golling, Tobias",
    title = "{CURTAINs for your Sliding Window: Constructing Unobserved Regions by Transforming Adjacent Intervals}",
    eprint = "2203.09470",
    archivePrefix = "arXiv",
    primaryClass = "hep-ph",
    month = "3",
    year = "2022"
}

@article{Hao_2023,
   title={Lorentz group equivariant autoencoders},
   volume={83},
   ISSN={1434-6052},
   url={http://dx.doi.org/10.1140/epjc/s10052-023-11633-5},
   DOI={10.1140/epjc/s10052-023-11633-5},
   number={6},
   journal={The European Physical Journal C},
   publisher={Springer Science and Business Media LLC},
   author={Hao, Zichun and Kansal, Raghav and Duarte, Javier and Chernyavskaya, Nadezda},
   year={2023},
   month=jun }

@article{Atkinson:2021nlt,
    author = "Atkinson, Oliver and Bhardwaj, Akanksha and Englert, Christoph and Ngairangbam, Vishal S. and Spannowsky, Michael",
    title = "{Anomaly detection with convolutional Graph Neural Networks}",
    eprint = "2105.07988",
    archivePrefix = "arXiv",
    primaryClass = "hep-ph",
    reportNumber = "IPPP/20/102",
    doi = "10.1007/JHEP08(2021)080",
    journal = "JHEP",
    volume = "08",
    pages = "080",
    year = "2021"
}

@article{Ngairangbam:2021yma,
    author = "Ngairangbam, Vishal S. and Spannowsky, Michael and Takeuchi, Michihisa",
    title = "{Anomaly detection in high-energy physics using a quantum autoencoder}",
    eprint = "2112.04958",
    archivePrefix = "arXiv",
    primaryClass = "hep-ph",
    reportNumber = "OU-HET-1125, IPPP/21/54",
    doi = "10.1103/PhysRevD.105.095004",
    journal = "Phys. Rev. D",
    volume = "105",
    number = "9",
    pages = "095004",
    year = "2022"
}

@article{Araz:2024lsl,
    author = "Araz, Jack Y. and Spannowsky, Michael",
    title = "{The role of data embedding in quantum autoencoders for improved anomaly detection}",
    eprint = "2409.04519",
    archivePrefix = "arXiv",
    primaryClass = "quant-ph",
    reportNumber = "JLAB-THY-24-4170, IPPP/24/60",
    month = "9",
    year = "2024"
}

@article{Bal:2025ydm,
    author = "Bal, Aritra and Klute, Markus and Maier, Benedikt and Oughton, Melik and Pezone, Eric and Spannowsky, Michael",
    title = "{1 Particle - 1 Qubit: Particle Physics Data Encoding for Quantum Machine Learning}",
    eprint = "2502.17301",
    archivePrefix = "arXiv",
    primaryClass = "hep-ph",
    reportNumber = "IPPP/25/11",
    month = "2",
    year = "2025"
}

@article{Finke:2021sdf,
    author = {Finke, Thorben and Kr\"amer, Michael and Morandini, Alessandro and M\"uck, Alexander and Oleksiyuk, Ivan},
    title = "{Autoencoders for unsupervised anomaly detection in high energy physics}",
    eprint = "2104.09051",
    archivePrefix = "arXiv",
    primaryClass = "hep-ph",
    reportNumber = "TTK-21-12",
    doi = "10.1007/JHEP06(2021)161",
    journal = "JHEP",
    volume = "06",
    pages = "161",
    year = "2021"
}

@inproceedings{Pol:2020weg,
    author = "Pol, Adrian Alan and Berger, Victor and Cerminara, Gianluca and Germain, Cecile and Pierini, Maurizio",
    title = "{Anomaly Detection With Conditional Variational Autoencoders}",
    booktitle = "{Eighteenth International Conference on Machine Learning and Applications}",
    eprint = "2010.05531",
    archivePrefix = "arXiv",
    primaryClass = "cs.LG",
    month = "10",
    year = "2020"
}

@article{Jawahar:2021vyu,
    author = "Jawahar, Pratik and Aarrestad, Thea and Chernyavskaya, Nadezda and Pierini, Maurizio and Wozniak, Kinga A. and Ngadiuba, Jennifer and Duarte, Javier and Tsan, Steven",
    title = "{Improving Variational Autoencoders for New Physics Detection at the LHC With Normalizing Flows}",
    eprint = "2110.08508",
    archivePrefix = "arXiv",
    primaryClass = "hep-ph",
    reportNumber = "FERMILAB-PUB-21-519-CMS",
    doi = "10.3389/fdata.2022.803685",
    journal = "Front. Big Data",
    volume = "5",
    pages = "803685",
    year = "2022"
}

@article{Canelli:2021aps,
    author = "Canelli, Florencia and de Cosa, Annapaola and Pottier, Luc Le and Niedziela, Jeremi and Pedro, Kevin and Pierini, Maurizio",
    title = "{Autoencoders for semivisible jet detection}",
    eprint = "2112.02864",
    archivePrefix = "arXiv",
    primaryClass = "hep-ph",
    doi = "10.1007/JHEP02(2022)074",
    journal = "JHEP",
    volume = "02",
    pages = "074",
    year = "2022"
}

@article{Atkinson:2022uzb,
    author = "Atkinson, Oliver and Bhardwaj, Akanksha and Englert, Christoph and Konar, Partha and Ngairangbam, Vishal S. and Spannowsky, Michael",
    title = "{IRC-Safe Graph Autoencoder for Unsupervised Anomaly Detection}",
    eprint = "2204.12231",
    archivePrefix = "arXiv",
    primaryClass = "hep-ph",
    doi = "10.3389/frai.2022.943135",
    journal = "Front. Artif. Intell.",
    volume = "5",
    pages = "943135",
    year = "2022"
}

@article{Chhibra:2023tyf,
    author = "Chhibra, Simranjit Singh and Chernyavskaya, Nadezda and Maier, Benedikt and Pierini, Maurzio and Hasan, Syed",
    title = "{Autoencoders for real-time SUEP detection}",
    eprint = "2306.13595",
    archivePrefix = "arXiv",
    primaryClass = "hep-ex",
    doi = "10.1140/epjp/s13360-024-05028-y",
    journal = "Eur. Phys. J. Plus",
    volume = "139",
    number = "3",
    pages = "281",
    year = "2024"
}

@article{Bradshaw:2022qev,
    author = "Bradshaw, Layne and Chang, Spencer and Ostdiek, Bryan",
    title = "{Creating simple, interpretable anomaly detectors for new physics in jet substructure}",
    eprint = "2203.01343",
    archivePrefix = "arXiv",
    primaryClass = "hep-ph",
    doi = "10.1103/PhysRevD.106.035014",
    journal = "Phys. Rev. D",
    volume = "106",
    number = "3",
    pages = "035014",
    year = "2022"
}

@article{Fraser:2021lxm,
    author = "Fraser, Katherine and Homiller, Samuel and Mishra, Rashmish K. and Ostdiek, Bryan and Schwartz, Matthew D.",
    title = "{Challenges for unsupervised anomaly detection in particle physics}",
    eprint = "2110.06948",
    archivePrefix = "arXiv",
    primaryClass = "hep-ph",
    doi = "10.1007/JHEP03(2022)066",
    journal = "JHEP",
    volume = "03",
    pages = "066",
    year = "2022"
}

@article{ANDERSON1972197,
title = {A simple neural network generating an interactive memory},
journal = {Mathematical Biosciences},
volume = {14},
number = {3},
pages = {197-220},
year = {1972},
issn = {0025-5564},
doi = {https://doi.org/10.1016/0025-5564(72)90075-2},
url = {https://www.sciencedirect.com/science/article/pii/0025556472900752},
author = {James A. Anderson}
}

@ARTICLE{6750072,
  author={Furber, Steve B. and Galluppi, Francesco and Temple, Steve and Plana, Luis A.},
  journal={Proceedings of the IEEE}, 
  title={The SpiNNaker Project}, 
  year={2014},
  volume={102},
  number={5},
  pages={652-665},
  doi={10.1109/JPROC.2014.2304638}}

@ARTICLE{2014Sci...345..668M,
       author = {{Merolla}, Paul A. and {Arthur}, John V. and {Alvarez-Icaza}, Rodrigo and {Cassidy}, Andrew S. and {Sawada}, Jun and {Akopyan}, Filipp and {Jackson}, Bryan L. and {Imam}, Nabil and {Guo}, Chen and {Nakamura}, Yutaka and {Brezzo}, Bernard and {Vo}, Ivan and {Esser}, Steven K. and {Appuswamy}, Rathinakumar and {Taba}, Brian and {Amir}, Arnon and {Flickner}, Myron D. and {Risk}, William P. and {Manohar}, Rajit and {Modha}, Dharmendra S.},
        title = "{A million spiking-neuron integrated circuit with a scalable communication network and interface}",
      journal = {Science},
         year = 2014,
        month = aug,
       volume = {345},
       number = {6197},
        pages = {668-673},
          doi = {10.1126/science.1254642},
       adsurl = {https://ui.adsabs.harvard.edu/abs/2014Sci...345..668M}
}

@ARTICLE{8259423,
  author={Davies, Mike and Srinivasa, Narayan and Lin, Tsung-Han and Chinya, Gautham and Cao, Yongqiang and Choday, Sri Harsha and Dimou, Georgios and Joshi, Prasad and Imam, Nabil and Jain, Shweta and Liao, Yuyun and Lin, Chit-Kwan and Lines, Andrew and Liu, Ruokun and Mathaikutty, Deepak and McCoy, Steven and Paul, Arnab and Tse, Jonathan and Venkataramanan, Guruguhanathan and Weng, Yi-Hsin and Wild, Andreas and Yang, Yoonseok and Wang, Hong},
  journal={IEEE Micro}, 
  title={Loihi: A Neuromorphic Manycore Processor with On-Chip Learning}, 
  year={2018},
  volume={38},
  number={1},
  pages={82-99},
  doi={10.1109/MM.2018.112130359}}

@INPROCEEDINGS{10448003,
  author={Shrestha, Sumit Bam and Timcheck, Jonathan and Frady, Paxon and Campos-Macias, Leobardo and Davies, Mike},
  booktitle={ICASSP 2024 - 2024 IEEE International Conference on Acoustics, Speech and Signal Processing (ICASSP)}, 
  title={Efficient Video and Audio Processing with Loihi 2}, 
  year={2024},
  volume={},
  number={},
  pages={13481-13485},
  doi={10.1109/ICASSP48485.2024.10448003}}

@misc{han2016deepcompressioncompressingdeep,
      title={Deep Compression: Compressing Deep Neural Networks with Pruning, Trained Quantization and Huffman Coding}, 
      author={Song Han and Huizi Mao and William J. Dally},
      year={2016},
      eprint={1510.00149},
      archivePrefix={arXiv},
      primaryClass={cs.CV},
      url={https://arxiv.org/abs/1510.00149}, 
}

@article{eshraghian2021training,
        title   =  {Training spiking neural networks using lessons from deep learning},
        author  =  {Eshraghian, Jason K and Ward, Max and Neftci, Emre and Wang, Xinxin
                    and Lenz, Gregor and Dwivedi, Girish and Bennamoun, Mohammed and
                   Jeong, Doo Seok and Lu, Wei D},
        journal = {Proceedings of the IEEE},
        volume  = {111},
        number  = {9},
        pages   = {1016--1054},
        year    = {2023}
}

@inproceedings{Ansel_PyTorch_2_Faster_2024,
author_ = {Ansel, Jason and Yang, Edward and He, Horace and Gimelshein, Natalia and Jain, Animesh and Voznesensky, Michael and Bao, Bin and Bell, Peter and Berard, David and Burovski, Evgeni and Chauhan, Geeta and Chourdia, Anjali and Constable, Will and Desmaison, Alban and DeVito, Zachary and Ellison, Elias and Feng, Will and Gong, Jiong and Gschwind, Michael and Hirsh, Brian and Huang, Sherlock and Kalambarkar, Kshiteej and Kirsch, Laurent and Lazos, Michael and Lezcano, Mario and Liang, Yanbo and Liang, Jason and Lu, Yinghai and Luk, CK and Maher, Bert and Pan, Yunjie and Puhrsch, Christian and Reso, Matthias and Saroufim, Mark and Siraichi, Marcos Yukio and Suk, Helen and Suo, Michael and Tillet, Phil and Wang, Eikan and Wang, Xiaodong and Wen, William and Zhang, Shunting and Zhao, Xu and Zhou, Keren and Zou, Richard and Mathews, Ajit and Chanan, Gregory and Wu, Peng and Chintala, Soumith},
author = {PyTroch Collaboration},
booktitle = {29th ACM International Conference on Architectural Support for Programming Languages and Operating Systems, Volume 2 (ASPLOS '24)},
doi = {10.1145/3620665.3640366},
month = apr,
publisher = {ACM},
title = {{PyTorch 2: Faster Machine Learning Through Dynamic Python Bytecode Transformation and Graph Compilation}},
url = {https://docs.pytorch.org/assets/pytorch2-2.pdf},
year = {2024}
}

@article{Deiana:2021niw,
    author = "Deiana, Allison McCarn and others",
    title = "{Applications and Techniques for Fast Machine Learning in Science}",
    eprint = "2110.13041",
    archivePrefix = "arXiv",
    primaryClass = "cs.LG",
    reportNumber = "FERMILAB-PUB-21-502-AD-E-SCD",
    doi = "10.3389/fdata.2022.787421",
    journal = "Front. Big Data",
    volume = "5",
    pages = "787421",
    year = "2022"
}

@article{bartlomiej_borzyszkowski_2020_3755310,
  author       = {Bartlomiej Borzyszkowski},
  title        = {Neuromorphic Computing in High Energy Physics},
  month        = apr,
  year         = 2020,
  publisher    = {Zenodo},
  doi          = {10.5281/zenodo.3755310},
  url          = {https://doi.org/10.5281/zenodo.3755310}
}

@ARTICLE{8891809,
  author={Neftci, Emre O. and Mostafa, Hesham and Zenke, Friedemann},
  journal={IEEE Signal Processing Magazine}, 
  title={Surrogate Gradient Learning in Spiking Neural Networks: Bringing the Power of Gradient-Based Optimization to Spiking Neural Networks}, 
  year={2019},
  volume={36},
  number={6},
  pages={51-63},
  doi={10.1109/MSP.2019.2931595}
}

@article{LI2022102765,
title = {Efficiency analysis of artificial vs. Spiking Neural Networks on FPGAs},
journal = {Journal of Systems Architecture},
volume = {133},
pages = {102765},
year = {2022},
issn = {1383-7621},
doi = {https://doi.org/10.1016/j.sysarc.2022.102765},
url = {https://www.sciencedirect.com/science/article/pii/S1383762122002508},
author = {Zhuoer Li and Edgar Lemaire and Nassim Abderrahmane and Sébastien Bilavarn and Benoit Miramond}
}

@article{KARAMIMANESH2025107256,
title = {Spiking neural networks on FPGA: A survey of methodologies and recent advancements},
journal = {Neural Networks},
volume = {186},
pages = {107256},
year = {2025},
issn = {0893-6080},
doi = {https://doi.org/10.1016/j.neunet.2025.107256},
url = {https://www.sciencedirect.com/science/article/pii/S0893608025001352},
}

@Article{Yao2024,
author={Yao, Man
and Richter, Ole
and Zhao, Guangshe
and Qiao, Ning
and Xing, Yannan
and Wang, Dingheng
and Hu, Tianxiang
and Fang, Wei
and Demirci, Tugba
and De Marchi, Michele
and Deng, Lei
and Yan, Tianyi
and Nielsen, Carsten
and Sheik, Sadique
and Wu, Chenxi
and Tian, Yonghong
and Xu, Bo
and Li, Guoqi},
title={Spike-based dynamic computing with asynchronous sensing-computing neuromorphic chip},
journal={Nature Communications},
year={2024},
month={May},
day={25},
volume={15},
number={1},
pages={4464},
issn={2041-1723},
doi={10.1038/s41467-024-47811-6},
url={https://doi.org/10.1038/s41467-024-47811-6},
}

@misc{plagwitz2023spikespikequantitativecomparison,
      title={To Spike or Not to Spike? A Quantitative Comparison of SNN and CNN FPGA Implementations}, 
      author={Patrick Plagwitz and Frank Hannig and Jürgen Teich and Oliver Keszocze},
      year={2023},
      eprint={2306.12742},
      archivePrefix={arXiv},
      primaryClass={cs.AR},
      url={https://arxiv.org/abs/2306.12742}, 
}

@article{rathi2023exploring,
  title={Exploring neuromorphic computing based on spiking neural networks: Algorithms to hardware},
  author={Rathi, Nitin and Chakraborty, Indranil and Kosta, Adarsh and Sengupta, Abhronil and Ankit, Aayush and Panda, Priyadarshini and Roy, Kaushik},
  journal={ACM Computing Surveys},
  volume={55},
  number={12},
  pages={1--49},
  year={2023},
  publisher={ACM New York, NY},
}

@article{TAVANAEI201947,
title = {Deep learning in spiking neural networks},
journal = {Neural Networks},
volume = {111},
pages = {47-63},
year = {2019},
issn = {0893-6080},
doi = {https://doi.org/10.1016/j.neunet.2018.12.002},
url = {https://www.sciencedirect.com/science/article/pii/S0893608018303332},
author = {Amirhossein Tavanaei and Masoud Ghodrati and Saeed Reza Kheradpisheh and Timothée Masquelier and Anthony Maida},
}

@misc{oconnor2016deepspikingnetworks,
      title={Deep Spiking Networks}, 
      author={Peter O'Connor and Max Welling},
      year={2016},
      eprint={1602.08323},
      archivePrefix={arXiv},
      primaryClass={cs.NE},
      url={https://arxiv.org/abs/1602.08323}, 
}

@article{Dillon:2025dxr,
    author = "Dillon, Barry M. and Spannowsky, Michael",
    title = "{Theory-informed neural networks for particle physics}",
    eprint = "2507.13447",
    archivePrefix = "arXiv",
    primaryClass = "hep-ph",
    reportNumber = "IPPP/25/47",
    month = "7",
    year = "2025"
}

\end{document}